\documentclass[aps,prl,reprint,preprintnumbers,superscriptaddress,nofootinbib,nobibnotes]{revtex4-2}
\usepackage[T1]{fontenc}
\usepackage{amsmath,amssymb,bm}
\usepackage{graphicx,booktabs}
\newcommand{\as}{\alpha_s}

\usepackage{blindtext}
\usepackage{lineno}
\usepackage{bm,color,xcolor}
\usepackage{slashed} 
\usepackage{graphicx}
\usepackage{subfigure} 
\usepackage{soul} 
\usepackage{amsmath}
\usepackage{multirow}
\usepackage{float}
\usepackage{comment}
\usepackage{fontawesome} 
\usepackage{mathtools}
\usepackage[colorlinks=true
,urlcolor=blue
,anchorcolor=blue
,citecolor=blue 
,filecolor=blue
,linkcolor=blue
,menucolor=blue
,linktocpage=true
,pdfproducer=medialab
,pdfa=true
]{hyperref}
\usepackage[nameinlink,capitalize]{cleveref} 
\usepackage{lipsum}

\makeatletter
\def\l@subsubsection#1#2{}
\makeatother

\newcommand{\GeV}{\ensuremath{\mathrm{GeV}}}

\newcommand{\beq}{\begin{equation}}
\newcommand{\eeq}{\end{equation}}
\newcommand{\bea}{\begin{eqnarray}}
\newcommand{\eea}{\end{eqnarray}}

\usepackage[arrows]{ezedits}
\definecolor{UMNRed}{RGB}{144,0,33}
\defineEditQuick{ZL}{UMNRed}

\begin{document}

\preprint{UMN-TH-4606/26}

\title{An Emergent Variance Scale in Learned Representations of QCD Jets}
\footnotetext{\textit{The authors are listed alphabetically.} Emails: \{\href{mailto:li001800@umn.edu}{li001800}, \href{mailto:zliuphys@umn.edu}{zliuphys}, \href{mailto:pan00389@umn.edu}{pan00389}, \href{mailto:nstrobbe@umn.edu}{nstrobbe}, \href{mailto:qianwen@umn.edu}{qianwen}\}@umn.edu}
\author{Peiran Li}
\affiliation{School of Physics and Astronomy, University of Minnesota, Minneapolis, MN 55455, USA}
\author{Zhen Liu}
\affiliation{School of Physics and Astronomy, University of Minnesota, Minneapolis, MN 55455, USA}
\author{Sitong Pan}
\affiliation{Department of Computer Science and Engineering, University of Minnesota, Minneapolis, MN 55455, USA}
\author{Nadja Strobbe}
\affiliation{School of Physics and Astronomy, University of Minnesota, Minneapolis, MN 55455, USA}
\author{Qianwen Wang}
\affiliation{Department of Computer Science and Engineering, University of Minnesota, Minneapolis, MN 55455, USA}
\date{September 26, 2026}

\begin{abstract}
We find an emergent common variance scale in variational autoencoders trained on QCD jets. The encoded mean features vary largely independently, in the sense of weak pairwise correlations. Surprisingly, the encoded variances change coherently from jet to jet, approximately rescaling by a single common factor. This collective organization emerges through unsupervised learning in both image-level and particle-level models trained on simulated light-quark and gluon jets. The common scale is sensitive to the underlying radiation dynamics: with the networks held fixed, the mean log variances respond coherently and approximately linearly to the strong coupling controlling final-state radiation in \textsc{Pythia}. This coupling produces the largest collective response among the shower and hadronization parameters examined over equal relative ranges. These findings reveal a simple, physically informative structure in the latent variances, offering an exciting opportunity to understand and interpret the representations learned by neural networks.
\end{abstract}
\maketitle
\raggedbottom

\begin{figure*}[!t]
 \centering
 \includegraphics[width=\textwidth,trim=3pt 150pt 13pt 118pt,clip]{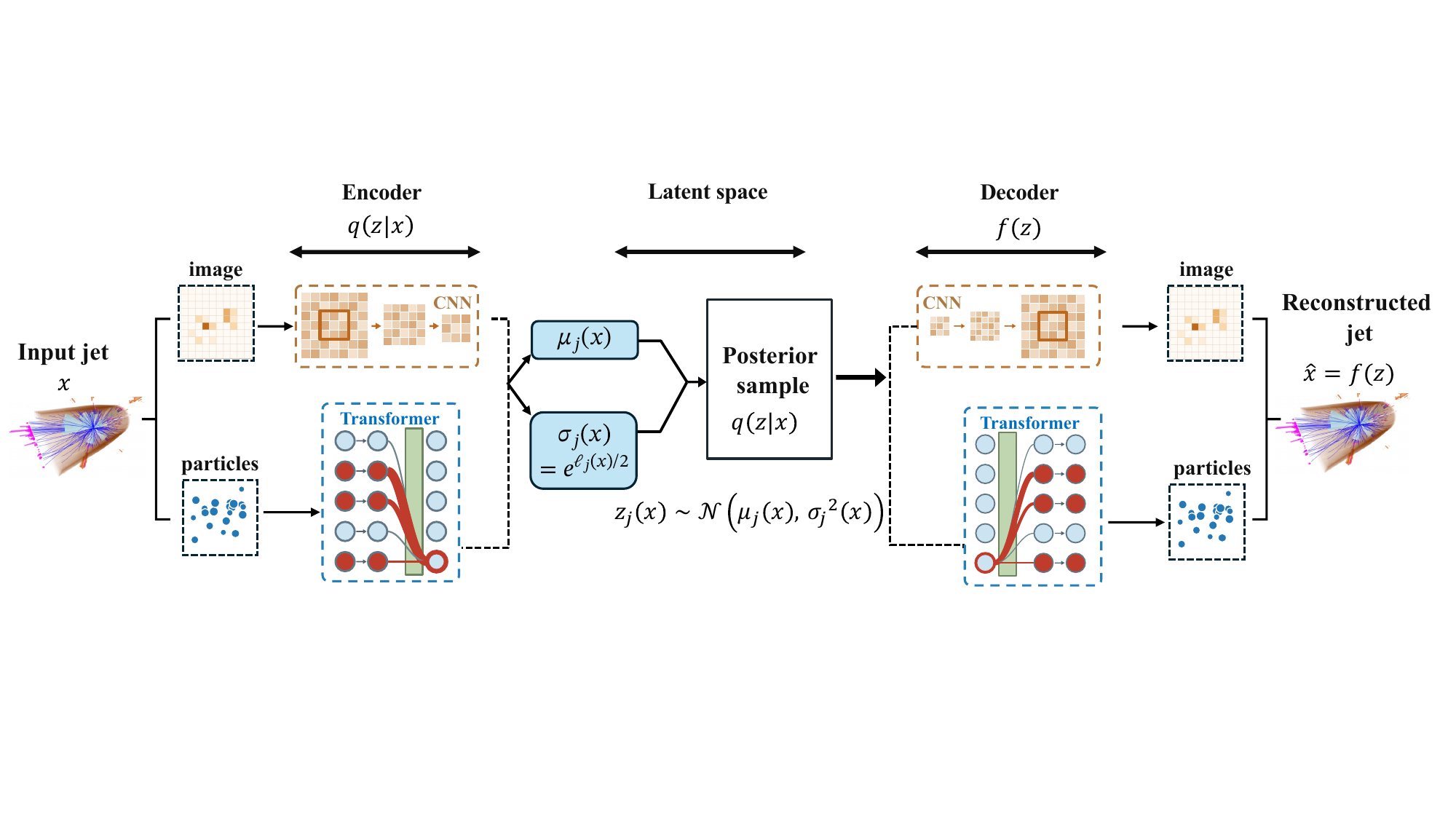}
 \caption{VAE architecture. The encoder maps a jet to the mean $\mu_j(x)$ and log variance $\ell_j(x)$ of a conditional Gaussian posterior. The latent coordinates $z_j$ are sampled from this posterior and passed to the decoder to reconstruct the jet. The learned widths $\sigma_j$ control the size of the random displacements from $\mu_j$.}
 \label{fig:vae_schematic}
\end{figure*}

\section{Introduction}

Machine learning (ML) has become a powerful tool for jet classification~\cite{deOliveira:2015xxd,Komiske:2018cqr}, anomaly detection~\cite{Heimel:2018mkt,Farina:2018fyg,Finke:2021sdf}, and simulation~\cite{Dohi:2020eda,Touranakou:2022qrp,Orzari:2023zrh}. Its success shows that neural networks can learn useful representations of the complex radiation patterns produced by quantum chromodynamics (QCD)~\cite{Larkoski:2017jix}. This raises a physical question: what information about jets do these representations retain, and how is it organized? Understanding this organization can help connect the features learned by a network to the dynamics of jet formation~\cite{Collins:2021pld,Faucett:2020vbu,Vent:2025ddm}. Our aim is to investigate such connections using a latent representation whose information content can be controlled through training.

Variational autoencoders (VAEs) provide a setting for this investigation~\cite{Kingma:2013hel}. A VAE reconstructs each input jet from a sampled latent state, balancing the reconstruction fidelity against an information cost imposed by the Kullback-Leibler (KL) divergence~\cite{Kullback:1951zyt} to a chosen prior. In a $\beta$-VAE, the relative weight $\beta$ controls this balance~\cite{DBLP:conf/iclr/HigginsMPBGBML17}. For a specified reconstruction loss and normalization, varying $\beta$ changes the level of jet detail retained by the representation and can change the number of active latent dimensions within a fixed network architecture. In this sense, $\beta$ provides control over the resolution of the learned representation~\cite{Collins:2021pld}. This makes it possible to study how the network allocates information among latent directions and with what precision it retains that information.

The learned representation contains both posterior means $\mu_j$, which locate each jet in latent space, and variances $\sigma_j^2$, which set the spread of the sampled latent coordinates around those means. We investigate how these two sets of quantities vary across jets and what their correlations reveal about the organization of jet information. Controlled changes to the simulated shower and hadronization dynamics then provide physical probes of the structures that emerge from training.

In this work, we train VAEs at the image- and particle-levels, using jet images and sets of jet constituents, respectively, on simulated quark and gluon jets generated with the default \textsc{Pythia} settings~\cite{Bierlich:2022pfr}. 
This setup allows us to draw conclusions that are not tied to a single representation choice.
Since training is unsupervised, the sensitivity of learned representations to generator parameters can reflect intrinsic jet structures rather than a supervised target.
We find a striking collective structure in the learned posterior parameters. In both image- and particle-level models, active features $\mu_j$ are generally weakly correlated, whereas the corresponding $\ln\sigma_j^2$ exhibit strong correlations and the $\sigma_j$ approximately share a common scaling law. We then hold the trained encoders fixed and independently vary parameters governing perturbative radiation and hadron formation. Among the parameters studied over equal fractional ranges, $\as$ produces the largest linear response in $\sigma$-space over the scanned range. This pronounced response appears in both the image- and particle-level models.

The common variance structure provides a first step toward understanding how these networks organize jet information, while its collective response to $\as$ probes the connection to radiation dynamics. Further interpretation of this structure may guide the construction of new, physically interpretable jet observables. Such observables could also be investigated as inputs to future determinations of $\as$. Establishing their practical usefulness and assessing their competitiveness with existing methods require further theoretical and experimental study.\footnote{Representative determinations use inclusive jet cross sections~\cite{CMS:2014qtp}, ratios of three-jet to two-jet rates~\cite{CMS:2013vbb}, event shapes~\cite{Abbate:2010xh}, and energy correlators~\cite{CMS:2024mlf}. Supervised neural-network approaches have also been developed to fit shower parameters, including $\as$, using complete particle-level information~\cite{Andreassen:2019nnm}.}

The paper is organized as follows. In \cref{sec:vae}, we introduce the VAE framework and the image-level and particle-level VAE models. In \cref{sec:collective}, we examine the correlations and common scaling behavior of the $\sigma$-space. We investigate its response to variations of the \textsc{Pythia} parameters and discuss a qualitative interpretation in terms of parton showering in \cref{sec:response}. Finally, in \cref{sec:outlook}, we summarize our findings and discuss possible applications and future directions. Additional comparisons across jet samples, representations and datasets (Appendices~\ref{app:zjet} and~\ref{app:datasets}), together with studies of multiplicity dependence (Appendix~\ref{app:rambo}) and correlations with standard jet observables (Appendix~\ref{app:observables}), further characterize the common variance structure and inform its physical interpretation.

\section{Variational jet representations}
\label{sec:vae}

A variational autoencoder (VAE), in this study, is a probabilistic generative model that encodes each jet as a latent distribution with a mean representation and widths that control sampling around it. Both components can carry information about jet structure. As shown in \cref{fig:vae_schematic}, the encoder maps the input jet to this latent distribution, and the decoder reconstructs the jet from a sampled latent point~\cite{Dohi:2020eda,Dillon:2021nxw}. Reconstruction encourages the latent representation to retain the input structure resolved by the chosen loss.

More precisely, for an input jet $x$, the encoder $q(x)$ returns two real values $\mu_j(x)$ and $\ell_j(x)=\ln\sigma_j^2(x)$ for each latent dimension $j$~\cite{Kingma:2013hel,Rezende:2014azh}. A latent point is drawn independently in each dimension,
\begin{align}
 z_j\sim\mathcal N\!\left(\mu_j(x),\sigma_j^2(x)\right),
 \label{eq:sampling}
\end{align}
and passed to the decoder $f(z)$. For the trained encoder, $\mu(x)$ and $\sigma(x)$ are fixed for a given jet, but $z$ is randomly sampled based on \cref{eq:sampling}. Repeated draws can therefore produce different reconstructions of the same input, with the widths $\sigma_j(x)$ controlling the random variation along each latent direction. 
Both the posterior means $\mu_j(x)$ and widths $\sigma_j(x)$ are learned through reconstruction and variational regularization. To test whether they encode information about QCD jet dynamics, we hold the trained encoder fixed and examine how their distribution responds to controlled variations of the generator parameters.

Ideally, the reconstructed jet would preserve all physical infrared- and collinear-safe features of the input without any over-training. Such a model would let us examine how jet structure is encoded and use the learned representation to investigate QCD dynamics. An excessively large latent space can allow the network to overfit by memorizing individual training jets. Such memorization alone offers little insight into the underlying physics. To encourage the networks to learn jet features, we therefore use compact latent spaces with 32 dimensions for both the image-level and particle-level VAEs, with $\mu_j$ and $\sigma_j$ for $j\in\{0,\ldots,31\}$. Perfect reconstruction is not required for this study. Capturing the main jet structure is sufficient to begin investigating the physical content of the latent representation.

\begin{figure}[!t]
 \centering
 \includegraphics[width=\columnwidth]{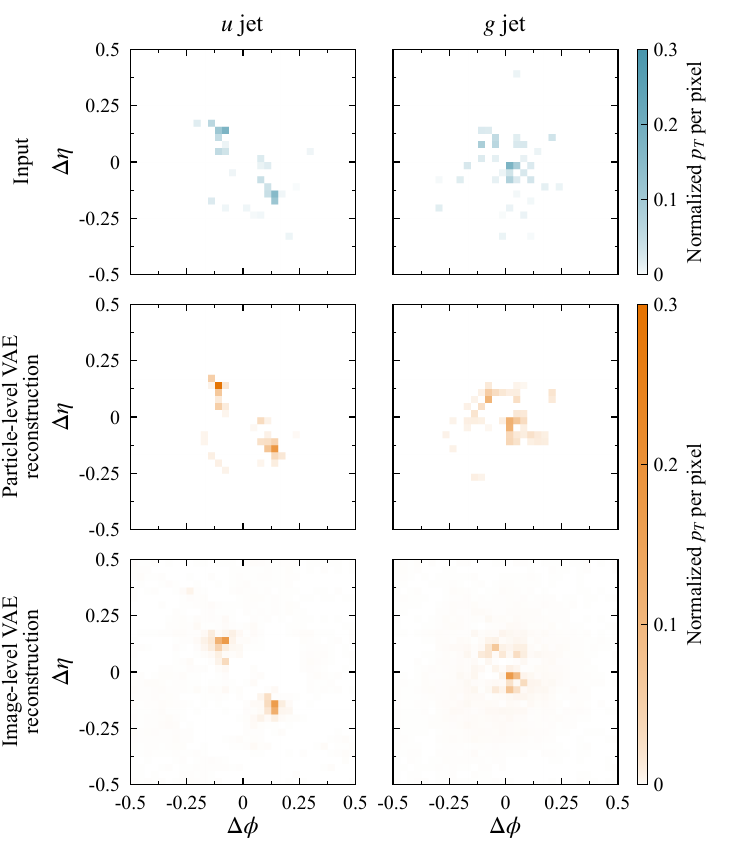}
 \caption{Illustrative reconstructions of one quark jet (left column) and one gluon jet (right column). The top row shows the inputs (blue); the middle and bottom rows show the particle-level and image-level VAE reconstructions (orange), respectively, both without PID inputs. All maps use $32\times32$ pixels over $-0.5\leq\Delta\eta,\Delta\phi\leq0.5$, with $\Delta\phi$ horizontal and $\Delta\eta$ vertical. The particle lists from the particle-level VAE are converted to the same image representation for display. All reconstructions are $f(\mu(x))$.}
 \label{fig:reconstruction}
\end{figure}

Training balances reconstruction against a standard Gaussian prior through the loss
\begin{align}
 L=L_{\mathrm{reco}}
   +\frac{\beta}{2}\sum_j\left[\mu_j^2+\sigma_j^2-\ln\sigma_j^2-1\right].
 \label{eq:loss}
\end{align}
Here $L_{\mathrm{reco}}$ measures the difference between the input and reconstructed jets as defined below. The second term is $\beta$ times the KL divergence from the posterior to the standard Gaussian prior~\cite{Kingma:2013hel,DBLP:conf/iclr/HigginsMPBGBML17}. It penalizes departures from zero means and unit variances, regularizing the latent representation. The relative weight $\beta$ controls the tradeoff between reconstruction and this regularization.

To ensure our insights are not tied to a single modeling choice, we consider two training scenarios that differ in input representation (a pixelated image vs. an unordered list of constituents), network architecture (CNN vs. Transformer), and reconstruction loss (pixel-wise squared error vs. a transport cost).
In \textit{image-level} training, a CNN-based VAE maps a normalized $32\times32$ transverse-momentum image to another image in the centered angular window $|\Delta\eta|,|\Delta\phi|<0.5$. The reconstruction loss $L_\text{reco}^\text{Image}$ is a squared-error cost summed over image pixels and averaged over jets. 
In \textit{particle-level} training, a VAE based on the Transformer architecture~\cite{Vaswani:2017lxt} takes a list of up to 80 jet constituents, represented by $(\hat{p}_T,\Delta\eta,\Delta\phi)$, where $\hat{p}_T=p_T/\sum_i p_{T,i}$ and the sum runs over the retained constituents. Its reconstruction loss $L_{\mathrm{reco}}^\text{Particle}$ is based on the energy mover's distance (EMD)~\cite{Komiske:2019fks,Collins:2021pld}, comparing the input and output $p_T$ distributions through a transport cost.\footnote{We use a debiased Sinkhorn transport cost~\cite{Feydy2019Sinkhorn}, an entropy-regularized approximation to the underlying optimal-transport cost.} Neither representation includes particle-identification (PID) information.

\cref{fig:reconstruction} illustrates both reconstruction schemes on example jets. Both models reproduce the main $p_T$ concentration and angular structure, while some diffuse radiation is redistributed. Appendix~\ref{app:analysis} discusses the training stability and reconstruction quality of both models in detail. 

\section{Collective posterior-width structure}
\label{sec:collective}

\subsection{Samples and reconstruction diagnostics}

A fixed dijet configuration provides a controlled starting point for studying the learned representations.
We simulate quark and gluon jet samples with \textsc{Pythia}~8.316~\cite{Bierlich:2022pfr} by showering and hadronizing manually initialized color-singlet parton pairs.\footnote{The example program \texttt{main234.cc}, distributed with \textsc{Pythia}~8.316, illustrates how to initialize parton configurations and generate their final-state showers and hadronization.} The default final-state shower and hadronization parameters follow the Monash 2013 tune~\cite{Skands:2014pea}. We initialize a color-singlet $u\bar u$ or $gg$ pair in its center-of-momentum frame, with an invariant mass of 400~GeV. This configuration has the same kinematics and color structure as the two-body decay of a 400~GeV Higgs-like particle at rest. The two partons each carry an energy of 200~GeV, with their three-momenta pointing along the $+x$ and $-x$ axes. We cluster the final-state particles with the anti-$k_T$ algorithm~\cite{Cacciari:2008gp}, using radius $R=0.4$, and select the highest-$p_T$ jet. The retained constituents are represented by their $(\hat{p}_T,\eta,\phi)$, without PID information. The resulting dataset contains 100\,000 quark jets and 100\,000 gluon jets for training and validation.

We train both the image-level and particle-level VAEs on the mixed sample of quark and gluon jets. \cref{fig:reconstruction} shows examples of the input jets and their reconstructions, obtained by decoding the posterior mean $\mu(x)$. These examples provide a qualitative check of the reconstructed jet structure and motivate examining how jet information is learned in the posterior means $\mu_j$ and posterior widths $\sigma_j$.



\subsection{Means and widths across events}

Correlations across jets probe relationships among the posterior means or widths of different latent dimensions. We evaluate these correlations for both the image-level and particle-level VAEs. \cref{fig:collective}(a,b,d,e) shows the Pearson correlation matrices among the active features $\mu_j$ and among the corresponding $\ln\sigma_j^2$ for both VAEs. We define a latent dimension $j$ as active when
\begin{align}
 \frac{\operatorname{Var}_x[\mu_j(x)]}{\mathbb{E}_x[\sigma_j^2(x)]}>1.
 \label{eq:activity}
\end{align}
This criterion selects dimensions for which the variance of the posterior mean across jets exceeds the average posterior sampling variance, so that variations between jets dominate over sampling fluctuations. For an active latent dimension $j$, we refer to $\mu_j(x)$ as an active feature; its corresponding posterior width $\sigma_j(x)$ controls sampling along that dimension. As shown in \cref{fig:collective}, the active features are generally weakly correlated, suggesting that they encode different jet information. However, most $\ln\sigma_j^2$ values are highly correlated, suggesting a dominant common underlying feature.

Previous studies have related learned jet representations to physical observables~\cite{Collins:2021pld,Vent:2025ddm}. Here we focus on the collective organization of the posterior widths, which control the spread around the mean representation but do not enter the decoder as additional inputs.


\subsection{Common scaling of \texorpdfstring{$\sigma$}{sigma}-space}

\begin{figure*}[t]
 \centering
 \includegraphics[width=\textwidth]{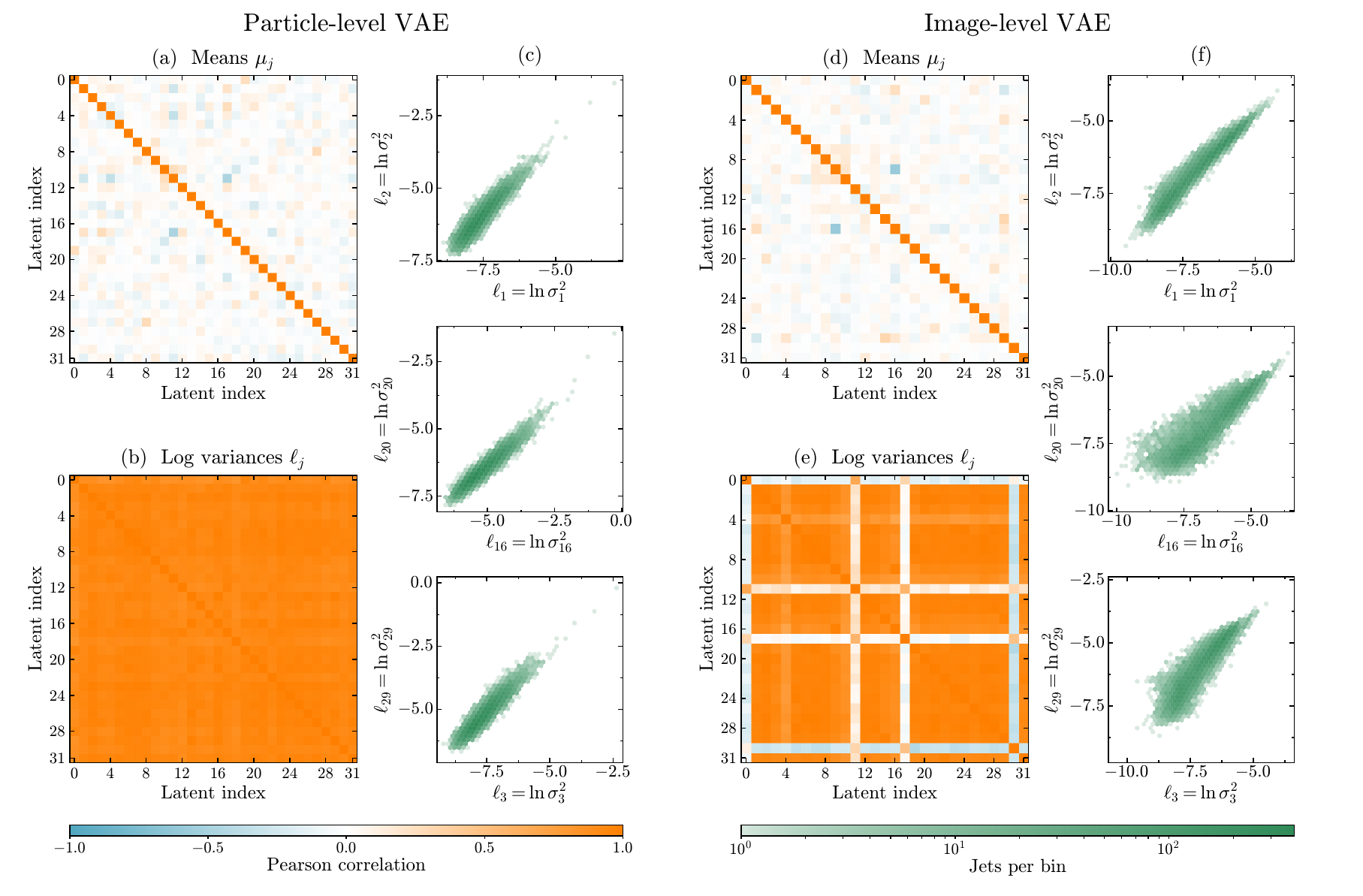}
 \caption{Posterior structure in the particle-level VAE with $\beta=10^{-5}$ (on the \textbf{left}) and the image-level VAE with $\beta=2\times10^{-6}$ (on the right). Panels \textbf{(a,d)} show the Pearson correlation matrices of the posterior means $\mu_j$, and panels \textbf{(b,e)} show those of the log variances $\ell_j=\ln\sigma_j^2$, restricted to the 32 and 29 active dimensions of the respective models as defined by~\cref{eq:activity}. Panels \textbf{(c)} and \textbf{(f)} each show the 2D joint distributions of log variances for three pairs of latent dimensions in the particle-level and image-level models, respectively. All panels use the same mixed sample of 5000 quark jets and 5000 gluon jets.}
 \label{fig:collective}
\end{figure*}

Strong correlations among the log variances suggest that much of their event-to-event variation can be represented by a common variable. 
We focus on the collective width scaling of the retained dimensions, described by
\begin{align}
 \ell_j(x)=\ln\sigma^2_j(x)\simeq a_j+b_j u(x),
 \label{eq:factor}
\end{align}
where $a_j,b_j$ are constants for each retained latent dimension and $u$ is a common event-dependent function for the fixed trained model. We normalize $u$ such that $\overline{b}=\sum_j b_j/N=1$, where the sum runs over the retained dimensions and $N$ is their number: 32 for the particle-level VAE and 25 for the image-level VAE.\footnote{In the image-level VAE, we exclude four active outliers, $j=0,11,17,30$ in \cref{fig:collective}(e), which do not follow the common log-variance trend. Their posterior variances ($\sigma_j^2\sim10^{-4}$) are much smaller than those of the remaining active dimensions (typically $\sigma_j^2\sim10^{-2}$). These coordinates are therefore nearly deterministic encodings of the input, $z_j\simeq\mu_j(x)$, with negligible fluctuations from posterior sampling. This work focuses on exploring the collective behavior of the posterior widths.}

The coefficients $b_j$ characterize how individual posterior widths scale with the common variable. Defining $U(x)=e^{u(x)/2}$, we obtain
\begin{align}
 \sigma_j(x)\simeq e^{a_j/2}
 \left(e^{u(x)/2}\right)^{b_j}
 =e^{a_j/2}U(x)^{b_j}.
 \label{eq:width_scaling}
\end{align}
Different values of $b_j$ therefore correspond to different scaling exponents. In general, the existence of a common variable $U(x)$ does not require these exponents to be equal. For the two VAEs shown in \cref{fig:collective}, the relative spreads over the retained dimensions are
\begin{align}
 \text{Particle-level}:~\frac{\sqrt{\sum_j\left[b_j-\overline{b}\right]^2/N}}{\overline{b}}
 &\simeq 0.0723,\label{eq:particle_loading_spread}\\
 \text{Image-level}:~\frac{\sqrt{\sum_j\left[b_j-\overline{b}\right]^2/N}}{\overline{b}}
 &\simeq 0.1020,\label{eq:image_loading_spread}
\end{align}
where the numerator is the standard deviation of the $b_j$ over the retained dimensions. Remarkably, these relative spreads of approximately $7\%$ and $10\%$ indicate that the scaling exponents cluster closely around unity.

The near-equal scaling exponents give an approximate common scaling law,
\begin{align}
 \sigma_j(x)\simeq e^{a_j/2}U(x).
\end{align}
Changing the input jet approximately rescales the posterior widths in the retained dimensions by a common factor, leaving their ratios approximately unchanged. The common scaling trend is illustrated by the joint distributions in \cref{fig:collective}(c,f), where the same three pairs of collective latent dimensions are shown for both models and their slopes are close to one. The weak correlations among the active features $\mu_j$ suggest that they encode different aspects of jet structure. Within the collective sector, the widths $\sigma_j$ are strongly correlated and approximately share the same scaling law.\footnote{In Appendix~\ref{app:datasets}, we compare the posterior log-variance structure in jets with that in MNIST (handwritten digits), Fashion-MNIST (clothing images), CelebA (face images), and dSprites (synthetic shapes). Correlations among posterior log variances also occur in these non-jet datasets, but their correlation and covariance matrices generally have higher effective ranks, indicating that the variation is less concentrated in a single collective mode. For jets, the effective ranks remain low across the tested KL weights $\beta$, even with substantial changes in the number of active dimensions.} 
The persistence of this collective structure in both VAEs motivates testing the sensitivity of the posterior widths to parameters governing jet formation.

\section{Coupling response and physical interpretation}
\label{sec:response}


The common variation of the $\sigma$-space raises the question of which aspects of jet formation are reflected in this behavior. We probe this dependence by varying the \textsc{Pythia} parameters governing perturbative radiation and hadron formation while holding the trained encoders fixed. 

\subsection{Pythia parameters vs. \texorpdfstring{$\sigma$}{sigma}-space}

A parameter scan measures how the width distribution changes when a specified part of the jet-generation model is varied. We examine eight parameters: the strong coupling $\as(m_Z^2)$, two Lund fragmentation parameters, the string transverse-momentum width, the final-state shower cutoff, and three hadron-species settings. Each parameter is varied independently by $-20\%,-10\%,0,+10\%,+20\%$ relative to its default value, while all other parameters are held fixed. For each parameter setting, we use \textsc{Pythia} 8.316~\cite{Bierlich:2022pfr} to shower and hadronize a color-connected, back-to-back $u\bar u$ or $gg$ pair with 200~GeV per parton. For every parameter benchmark, we generate separate quark and gluon samples, each containing 10\,000 jets, and combine them into an equally weighted sample of 20\,000 jets for the response analysis.

Here, $\as(m_Z^2)$ sets the strength of final-state radiation, while $p_{T,\min}$ specifies the infrared (IR) cutoff on the transverse-momentum evolution variable of the final-state shower. The Lund parameters $a_{\rm Lund}$ and $b_{\rm Lund}$ govern the longitudinal momentum sharing during string fragmentation, and $\sigma_{\rm string}$ sets the transverse-momentum width generated at string breaking~\cite{Andersson:1983ia}. The hadron-species parameters $r_{V/P}$, $r_{s/(u,d)}$, and $r_{qq/q}$ control, respectively, the vector-to-pseudoscalar production ratio for light mesons, strange-quark suppression relative to $u$ or $d$, and diquark suppression relative to quarks, which affects baryon production~\cite{Bierlich:2022pfr}. The corresponding \textsc{Pythia} settings, default values, and scan ranges are given in Appendix~\ref{app:analysis}.

For each parameter value $p$, we pass the mixed $u/g$-jet sample through the above trained VAE while keeping its encoder fixed.\footnote{Both image- and particle-level VAEs are trained only on samples generated with the default \textsc{Pythia} parameters $p_0$.} Averaging over the jets gives a list of mean log variances ${\langle\ell_j\rangle_p}$, where $j$ labels the active latent dimensions, $\langle\cdots\rangle\equiv\mathbb{E}_x[\cdots]$. To study their parameter dependence, we compare these means with those obtained at the default value $p_0$,
\begin{align}
 \Delta \langle \ell_j\rangle(p)=\langle\ell_j\rangle_p-\langle\ell_j\rangle_{p_0}.
 \label{eq:widthshift}
\end{align}
We do not further normalize this difference by $\langle\ell_j\rangle_{p_0}$, since $\Delta\langle\ell_j\rangle(p)$ already measures relative scaling: 
\begin{align}
    \Delta\langle\ell_j\rangle(p)
    = \ln\frac{\exp\langle\ln\sigma_j^2\rangle_p}{\exp\langle\ln\sigma_j^2\rangle_{p_0}}.
\end{align}

\begin{figure*}[t]
 \centering
 \includegraphics[width=\textwidth]{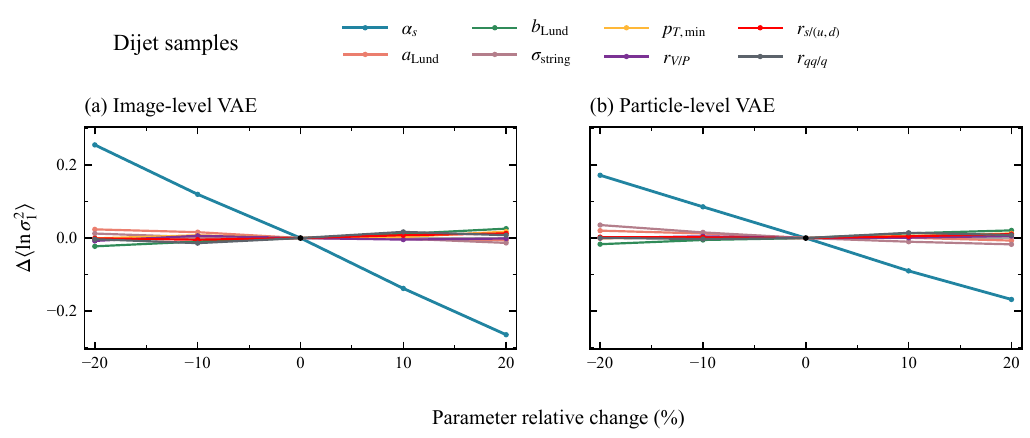}
 \caption{Parameter responses of an active dimension, $j=1$, in \textbf{(a)} the image-level VAE and \textbf{(b)} the particle-level VAE. Each sample combines 10\,000 $u$ jets and 10\,000 gluon jets from the dijet setup. The eight curves show $\Delta\langle\ln\sigma_1^2\rangle$ relative to the default settings. Small colored dots mark the four nonzero parameter variations; a single black dot marks the shared default point. Lines connect the five measured points.}
 \label{fig:alpha}
\end{figure*}

\cref{fig:alpha} shows the response of an active dimension, $j=1$, to the eight parameter variations. The change in $\langle\ell_j\rangle_p$ is much larger for $\as$ than for the other seven parameters and is approximately linear over the scanned range. This behavior appears in both the image-level and particle-level VAEs.

\begin{figure*}[t]
 \centering
 \includegraphics[width=0.9\textwidth]{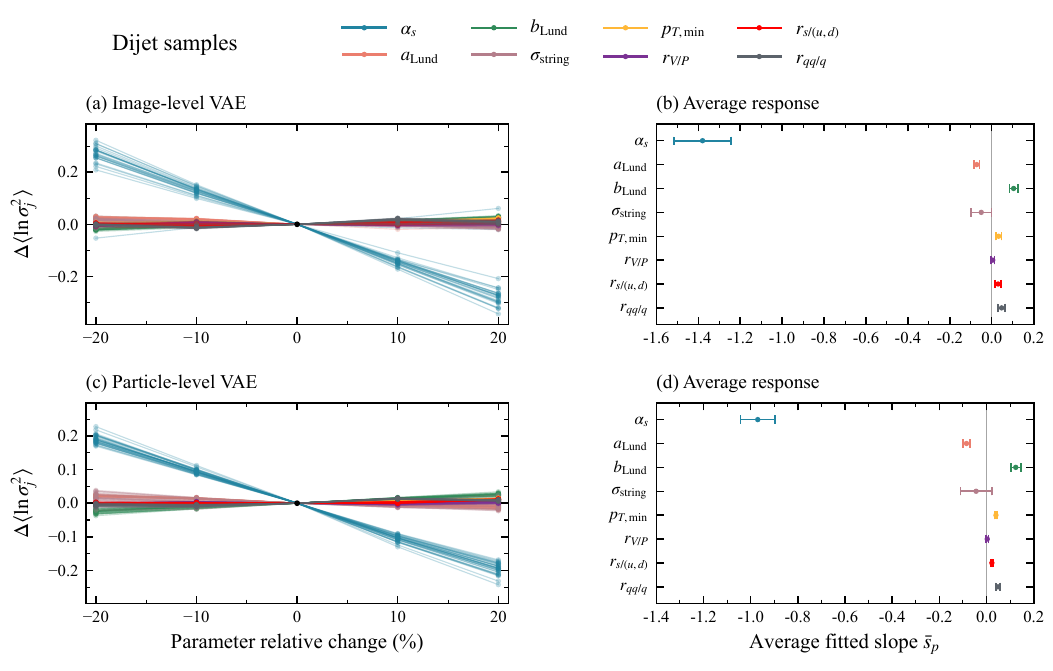}
 \caption{Collective parameter responses in the image-level VAE (top row) and particle-level VAE (bottom row). \textbf{(a,c)} Each thin curve connects the five measured shifts for one active dimension and one generator parameter. All 29 active dimensions of the image-level VAE and all 32 of the particle-level VAE are shown. \textbf{(b,d)} Small dots give the average fitted slopes $s_{j,p}$ over the active dimensions. Horizontal error bars show the standard deviation of the fitted slope across these dimensions. Four active dimensions in the image-level VAE are clear outliers, exhibiting much weaker responses to $\as$ than the others. To focus on the collective behavior, these four dimensions are excluded from the $\as$ average and standard deviation in (b).}
 \label{fig:parameters}
\end{figure*}

We next examine whether this response is shared across the active dimensions. Applying~\cref{eq:activity} to the $u/g$-mixed sample at the default settings $p_0$ identifies 29 active dimensions in the image-level VAE and 32 in the particle-level VAE; each set is kept fixed throughout the scans. The left panels of \cref{fig:parameters} show the response curves for every active dimension and all eight parameters. For each active dimension and parameter, we fit
\begin{align}
 \Delta\langle\ell_j\rangle(p)\simeq c_{j,p}+s_{j,p}\frac{p-p_0}{p_0},
 \label{eq:response}
\end{align}
with a free intercept $c_{j,p}$ and slope $s_{j,p}$. 
The right panels summarize the fitted slopes, with error bars indicating dispersion across dimensions. 
$\as$ gives the largest absolute fitted slope for most of the active dimensions in both models. Thus, the strong, approximately linear response to the \textsc{Pythia} coupling parameter is a collective property of $\sigma$-space in both the image-level and particle-level VAEs.

An additional cross-check using $pp\to Z(\to\nu\bar{\nu})+\mathrm{jet}$ samples shows the same qualitative hierarchy of parameter responses; see Appendix~\ref{app:zjet}.

\subsection{A qualitative interpretation from parton showering}


Parton showers describe radiation as a sequence of emissions, governed by splitting and no-emission probabilities. Changing the shower coupling modifies radiation probabilities and, through the generator's infrared prescription, can also shift the effective cutoff.

At one loop, the running coupling defines an IR scale $\Lambda_{n_f}$ through~\cite{Skands:2012ts}
\begin{align}
 \Lambda_{n_f}^2
 =\mu^2\exp\!\left[-\frac{1}{b_0\as(\mu^2)}\right],
 \label{eq:showerlambda}
\end{align}
where $b_0=(33-2n_f)/(12\pi)$ and $n_f$ is fixed within each flavor interval. Matching across quark thresholds determines the three-light-flavor scale $\Lambda_3$. With the settings used in \textsc{Pythia}, $\Lambda_3$ enters the IR cutoff on the transverse-momentum evolution variable through $k_{T,\min}=\max\{p_{T,\min},1.1\Lambda_3\}$~\cite{Bierlich:2022pfr}, where the default value of $p_{T,\min}$ is 0.5~GeV. Therefore, varying the input $\as(m_Z^2)$ can change the effective IR cutoff.

Changing the IR cutoff alone cannot explain the observed response: lowering $\as$ leaves the effective cutoff fixed at 0.5~GeV, yet $\sigma$-space still responds strongly. The independent scan of \texttt{TimeShower:pTmin} over 0.4--0.6~GeV also produces a weaker response in $\sigma$-space than the $\as$ scan. For the $+10\%$ and $+20\%$ variations of $\as$, the effective cutoff rises to approximately 0.569 and 0.838~GeV, respectively. Although the latter lies beyond the independent cutoff scan, the $+10\%$ variation of $\as$ already produces a substantially stronger response in $\sigma$-space than increasing $p_{T,\min}$ to 0.6~GeV at fixed $\as$.

Varying $\as$ also modifies the splitting probabilities. For a given starting state at scale $Q$, the exclusive differential probability for the next splitting in a transverse-momentum-ordered shower is schematically~\cite{Bierlich:2022pfr,Lonnblad:2012hz}
\begin{align}
 \mathrm{d}\mathcal P^{\mathrm{next}}_{a\to bc}
 \simeq{}&\Delta_a(Q^2,k_T^2)\,
 \frac{\as(k_T^2)}{2\pi}
 \, P_{a\to bc}(z)
 \frac{\mathrm{d}k_T^2}{k_T^2}\,\mathrm{d}z,
 \label{eq:nextbranching}
\end{align}
where $k_T$ is the shower's transverse-momentum evolution variable, $z$ is the daughter momentum fraction, and $P_{a\to bc}(z)$ is the splitting function~\cite{Altarelli:1977zs}. The factors following $\Delta_a$ give the local differential splitting rate; $\Delta_a$ accounts for the absence of any preceding splitting between $Q$ and $k_T$. Increasing the input $\as(m_Z^2)$ raises the running $\as(k_T^2)$ across the common perturbative range and hence increases the local splitting rate at fixed kinematics.

The corresponding probability for no splitting between the starting scale $Q$ and a lower scale $k_T$ is the Sudakov factor,
\begin{align}
 \Delta_a(Q^2,k_T^2)
 ={}&\exp\!\Bigg[-\sum_{b,c}
 \int_{k_T^2}^{Q^2}\frac{\mathrm{d}\kappa^2}{\kappa^2}
 \nonumber\\
 &\times
 \int_{z_{\min}(\kappa^2)}^{z_{\max}(\kappa^2)}
 \mathrm{d}z'\,\frac{\as(\kappa^2)}{2\pi}
 P_{a\to bc}(z')\Bigg].
 \label{eq:sudakov}
\end{align}
For a fixed starting state and common phase space above the cutoff, a larger $\as$ decreases $\Delta_a$. The probability for a splitting before the evolution reaches $k_T$ is $1-\Delta_a(Q^2,k_T^2)$ and therefore increases. In this sense, the next splitting is more likely to occur at a higher evolution scale. Through successive splittings, the modified emission and no-emission probabilities systematically change the energy sharing and angular distribution of the final jet.

The strong, approximately linear response to $\as$ in \cref{fig:alpha,fig:parameters} offers an important clue: information about jet structure shaped by stochastic shower dynamics may be encoded in $\sigma$-space during training. Varying $\as$, which governs these jet dynamics, systematically changes the statistical distribution of the posterior widths $\sigma_j$. Together with the strong correlations and approximate common scaling observed in~\cref{sec:collective}, these results support the interpretation that the posterior widths $\sigma_j$ associated with the active dimensions encode common information about the underlying radiation dynamics.

The effective rank of the log-variance space characterizes how many linear modes contribute appreciably to the variation of $\ln\sigma_j^2$ across inputs. Multiple modes can describe distinct patterns of uncertainty variation among latent features. The near-one rank observed for jets instead indicates predominantly collective variation, even though the mean representation retains many effective directions; the cross-dataset comparison in Appendix~\ref{app:datasets} develops this interpretation.

Correlations with standard jet observables provide complementary information about the learned representation. As shown in Appendix~\ref{app:observables}, the representative log variance $\ell_0$ is correlated with constituent multiplicity, momentum sharing and angular structure, while several mean coordinates also exhibit substantial correlations with these observables. These associations connect the learned variance scale to familiar properties of QCD jets. Specifically, we compare QCD jets with uniform phase-space samples in Appendix~\ref{app:rambo} and show that particle multiplicity alone is insufficient to explain the collective behavior of $\sigma$-space in the particle-level VAE.

These observations establish an empirical connection between the learned widths and coupling-dependent jet structure. Pairwise correlations do not determine whether a single observable or a combination of observables accounts for the common mode. Nor do they explain why the log variances vary approximately together across active dimensions and exhibit a coherent coupling response. A quantitative explanation of this collective behavior in terms of the reconstruction objective and QCD dynamics remains open.

\section{Conclusions and Outlook}
\label{sec:outlook}

In this work, we have explored the physical information encoded in the conditional sampling distributions of variational jet representations. Networks trained through jet reconstruction develop strongly correlated posterior widths with an approximate common scaling behavior. With the trained encoders held fixed, the mean log variances exhibit a coherent, approximately linear response to the strong coupling in both image-level and particle-level representations. Among the \textsc{Pythia} parameters examined, the strong coupling produces the largest response in the controlled dijet study. These observations provide evidence that the sampling structure of a learned representation retains information about the QCD dynamics underlying jet formation.

One possible application is a determination of $\as$ using statistics derived from the learned sampling distributions. The sensitivity identified here emerges without generator-parameter labels or a parameter-estimation objective during training. This differs from approaches in which machine learning is explicitly used to optimize generator parameters~\cite{Lazzarin:2020uvv,Andreassen:2019nnm,Ilten:2016csi}. With the encoder held fixed, the observed coupling dependence of $\sigma$-space could provide a basis for a future determination of $\as$ from jet data. Establishing the quantitative usefulness of these statistics would require cross-checks with jets from different production processes and channels to test the robustness of the observed coupling response. Applications to real data at the LHC would also require assessing detector effects, experimental uncertainties, and generator-model dependence.

The implications of this study extend beyond sensitivity to a single generator parameter. Varying the strong coupling modifies both local splitting rates and Sudakov factors, motivating a broader investigation of the radiation dynamics encoded in latent sampling distributions. Controlled variations of splitting kernels, accompanied by consistent changes in the Sudakov factors~\cite{Mrenna:2016sih}, could test whether these distributions distinguish changes in the overall amount of radiation from changes in its energy-sharing and angular structure. Complementary studies using jets at different energies and angular resolutions could explore connections to the running coupling, the QCD scale $\Lambda_{\mathrm{QCD}}$, and renormalization-group evolution, with shower-cutoff and hadronization effects carefully accounted for.

Although we use Gaussian conditional distributions in this study, the broader idea does not require a Gaussian form. More general distributions would allow information to be carried by their shape and correlations as well as their widths~\cite{Rezende:2015ocs}.
Our results demonstrate the potential of the distributions used to sample latent features as an additional source of physical information, motivating further exploration of latent sampling distributions as probes of the QCD dynamics underlying jet formation.

\section{Acknowledgments}
We thank Benjamin P. Nachman, David Shih and Jesse Thaler for helpful discussions. Z.L. thanks Jack Collins, Ke-Pan Xie, Yiheng Jiang, Lian-Tao Wang, and Yingying Li for early discussions and collaborations exploring variational autoencoders in jets.

This work was supported in part by the GenAI4Science seed program. We acknowledge the CSE Data Science Initiative at the University of Minnesota College of Science and Engineering for funding through the ADC Graduate Assistantship. P.L. and Z.L. are supported in part by the U.S. Department of Energy under Grant No.~DE-SC0011842 at the University of Minnesota. N.S. is supported by the U.S. Department of Energy, Office of Science, Office of High Energy Physics under Award No.~DE-SC0011845. P.L. is also supported in part by a Doctoral Dissertation Fellowship at the University of Minnesota.

We acknowledge the Minnesota Supercomputing Institute (MSI) at the University of Minnesota for providing computational resources used in this work.

Code and numerical data accompanying this study are publicly available at \href{https://github.com/MLQCD/jet-vae}{GitHub~\faGithub}.\\

\bibliography{references}

\onecolumngrid
\appendix

\section{Samples and analysis details}
\label{app:analysis}

This appendix provides details of the jet samples and preprocessing, the VAE training setup and reconstruction quality, and the generator parameters and scan ranges used in the analysis.

\subsection{Samples, preprocessing, and training}
The generated final state and constituent preprocessing together determine the jet information available to the encoder. The dijet sources contain complete final states from manually initialized color-connected parton pairs. Final-state showering is invoked with a maximum scale of 200~GeV. The analysis clusters these states using the eta-distance implementation of anti-$k_T$, $R=0.4$, and retains the leading jet with $p_T>10$~GeV. The $Z+$jet sample is described in Appendix~\ref{app:zjet}.

The image-level VAE uses the normalized total-$p_T$ image channel in the window $|\Delta\eta|,|\Delta\phi|<0.5$. For the particle-level VAE, constituents are ordered by $p_T$, the first 80 are retained, their fractions are renormalized, and the summed momentum defines the centered angular coordinates. Shorter lists are padded and masked. No PID input is used. The posterior analyses of both VAEs in \cref{fig:collective} use the same balanced sample of 5000 quark jets and 5000 gluon jets randomly selected from the default samples used for training and validation.

Both VAEs have 32 latent dimensions. The image-level VAE uses logarithmic annealing of the effective KL coefficient in \cref{eq:loss},
\begin{equation}
 \beta_{\rm img}(e)=\beta_{\rm init}
 \left(\frac{\beta_{\rm final}}{\beta_{\rm init}}\right)^{\min(e/150,1)},
 \label{eq:image-beta-annealing}
\end{equation}
where $e$ is the training epoch, $\beta_{\rm init}=1$, and $\beta_{\rm final}=2\times10^{-6}$. Thus $\ln\beta_{\rm img}$ decreases linearly over approximately 150 epochs, after which the coefficient remains fixed at $2\times10^{-6}$. The image-level results in \cref{fig:reconstruction,fig:collective,fig:alpha,fig:parameters} use the checkpoint selected at epoch 739 from this 800-epoch training run.

With this effective KL normalization, the image-level reconstruction loss is
\begin{equation}
 L_{\mathrm{reco}}^\text{Image}
 =\sum_{p=1}^{32\times 32}
 \left(x_p-\tilde{x}_p\right)^2,
 \label{eq:image-reconstruction-loss}
\end{equation}
where $p$ labels the pixels of the single-channel $32\times32$ image. The input and reconstructed pixel values are normalized transverse-momentum fractions, satisfying $\sum_p x_p=\sum_p\tilde{x}_p=1$ for each jet. During training, this per-jet reconstruction loss is averaged over each batch of 256 jets. The pixel contributions are summed without division by the number of pixels.

The particle-level VAE uses a debiased Sinkhorn approximation to the energy mover's distance~\cite{Komiske:2019fks,Feydy2019Sinkhorn}. For an input jet $x$ and its reconstruction $\tilde x$, let $a_i$ and $b_k$ denote the respective constituent $p_T$ fractions, normalized such that $\sum_i a_i=\sum_k b_k=1$. The entropically regularized transport cost is
\begin{equation}
 \begin{aligned}
 \mathcal T_\epsilon(x,\tilde x)
 =\min_{T\in\Pi(a,b)}\bigg\{&\sum_{ik}T_{ik}\frac{\Delta R_{ik}}{R}+\epsilon\sum_{ik}T_{ik}\ln\frac{T_{ik}}{a_i b_k}\bigg\},
 \end{aligned}
 \label{eq:regularized-transport}
\end{equation}
where $R=0.4$ and $\Delta R_{ik}=\sqrt{(\eta_i-\tilde\eta_k)^2+(\Delta\phi_{ik})^2}$, with $\Delta\phi_{ik}$ taken modulo $2\pi$ in $[-\pi,\pi]$. $T_{ik}$ denotes the fraction of transverse momentum transported from constituent $i$ of the input jet to constituent $k$ of the reconstructed jet. The set $\Pi(a,b)$ contains transport matrices with row sums $a_i$ and column sums $b_k$; zero-weight constituents are omitted. The minimization is approximated using Sinkhorn iterations~\cite{NIPS2013_af21d0c9}. The two self-transport costs are subtracted to obtain the Sinkhorn divergence,
\begin{align}
 S_\epsilon(x,\tilde x)={}&\mathcal T_\epsilon(x,\tilde x)
 -\tfrac12\mathcal T_\epsilon(x,x)-\tfrac12\mathcal T_\epsilon(\tilde x,\tilde x).
 \label{eq:sinkhorn-divergence}
\end{align}
The reconstruction loss $L_{\mathrm{reco}}$ is the batch average of $\max\{S_\epsilon(x,\tilde x),0\}$, with the clipping enforcing nonnegativity of the numerical estimate. The entropic regularization parameter follows
\begin{equation}
 \epsilon(e)=0.05\left(\frac{0.01}{0.05}\right)^{\min(e/50,1)},
 \label{eq:sinkhorn-annealing}
\end{equation}
where $e$ is the training epoch. It decreases geometrically to 0.01 over the first 50 epochs and then remains fixed, while the particle-level KL coefficient remains $\beta=10^{-5}$ throughout training.

\subsection{Reconstruction quality}
\label{app:reconstruction}

\begin{figure}[htbp]
 \centering
 \includegraphics[width=0.48\textwidth]{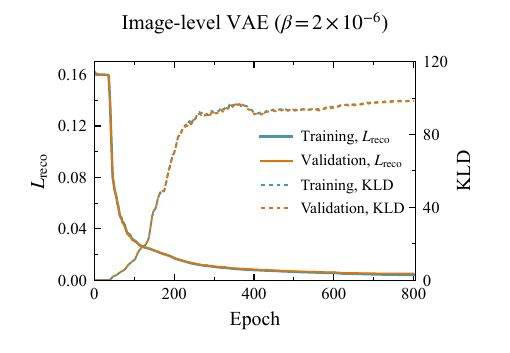}\hfill
 \includegraphics[width=0.48\textwidth]{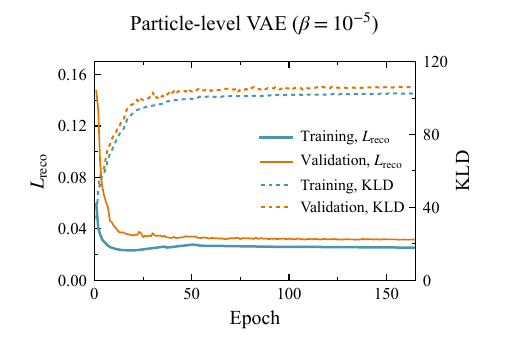}
 \caption{Training evolution of the image-level VAE with $\beta=2\times10^{-6}$ \textbf{(left)} and the particle-level VAE with $\beta=10^{-5}$ \textbf{(right)}. In each panel, solid curves show the reconstruction loss $L_{\mathrm{reco}}$ on the left axis, and dashed curves show the KL divergence (not multiplied by $\beta$) on the right axis.}
 \label{fig:cnn_training}
\end{figure}

The reconstruction losses decrease during training and approach approximately stable values at late epochs. For the image-level VAE with $\beta=2\times10^{-6}$, the training and validation curves closely track each other, as shown in \cref{fig:cnn_training}. These curves are approximately digitized from the saved training plot and illustrate the overall training trends. For the particle-level VAE, the validation loss remains above the training loss, with an approximately stable gap after the Sinkhorn regularization schedule is completed. These trends are consistent with stabilization of the optimization. The numerical values of $L_{\mathrm{reco}}$ are not directly comparable between the two VAEs, because the image-level model uses an MSE-based loss whereas the particle-level model uses an EMD-based transport loss. The image-level loss squares differences between normalized pixel $p_T$ fractions, whereas the particle-level EMD cost weights transport distances linearly by the transported $p_T$ fractions. This difference contributes to the typically smaller numerical values of the image-level $L_{\mathrm{reco}}$.

The examples in \cref{fig:reconstruction} illustrate the retention of the dominant transverse-momentum concentration and angular structure, together with some redistribution of diffuse radiation. In particular, the particle-level EMD loss function is IR-safe and only compares the normalized angular distributions of transverse momentum without requiring a one-to-one correspondence between input and reconstructed constituents. Constituent multiplicities and individual momentum fractions therefore need not be reproduced in particle-level VAE.

\subsection{Generator directions}
The dijet $\as$ scan varies \texttt{TimeShower:alphaSvalue}, with default value 0.1365 and range 0.1092--0.1638. The parameters \texttt{aLund} and \texttt{bLund} in \texttt{StringZ} determine the shape of the Lund symmetric fragmentation function, which governs longitudinal momentum sharing during string fragmentation. The parameter \texttt{sigma} in \texttt{StringPT} sets the width of the transverse-momentum kicks generated at string breaking. The input infrared cutoff on the transverse-momentum evolution variable of the final-state parton shower is specified by \texttt{TimeShower:pTmin}. The final three rows of \cref{tab:ranges} correspond to \texttt{mesonUDvector}, \texttt{probStoUD}, and \texttt{probQQtoQ} in \texttt{StringFlav}. These control, respectively, the vector-to-pseudoscalar production ratio for light ($u,d$) mesons, the suppression of strange-quark production relative to $u$ or $d$ quarks, and the suppression of diquark production relative to quark production, which regulates baryon production. The last three denote generator relative production settings, not normalized exclusive probabilities. The joint $Z+$jet coupling variation is specified in Appendix~\ref{app:zjet}. At the positive dijet $\as$ variations, the generator raises its internal effective shower cutoff from 0.5~GeV to approximately 0.569 and 0.838~GeV. The displayed response (\cref{fig:alpha,fig:parameters}) includes this implementation effect.

\begin{table}[H]
 \caption{Default inputs and scan ranges. Each axis uses the same five fractional steps. The coupling row gives the shower inputs; the $Z+$jet hard and MPI inputs are specified in the text. $b_{\rm Lund}$ is in $\GeV^{-2}$, and the two transverse-momentum scales are in GeV.}
 \label{tab:ranges}
 \begin{ruledtabular}
 \begin{tabular}{lcc}
 Parameter & Default & Range\\
 $\as(m_Z^2)$ & 0.1365 & 0.1092--0.1638\\
 $a_{\rm Lund}$ & 0.68 & 0.544--0.816\\
 $b_{\rm Lund}$ & 0.98 & 0.784--1.176\\
 $\sigma_{\rm string}$ & 0.335 & 0.268--0.402\\
 $p_{T,\min}$ & 0.5 & 0.4--0.6\\
 $r_{V/P}$ & 0.5 & 0.4--0.6\\
 $r_{s/(u,d)}$ & 0.217 & 0.1736--0.2604\\
 $r_{qq/q}$ & 0.081 & 0.0648--0.0972
 \end{tabular}
 \end{ruledtabular}
\end{table}

\section{Cross-check with \texorpdfstring{$Z+$jet}{Z+jet} production}
\label{app:zjet}

To test the fixed encoders beyond the dijet sample setup, we use jets produced in association with an invisibly decaying $Z$ boson. This process provides a widely used quark/gluon benchmark, including in studies of Energy Flow Networks and ParticleNet~\cite{Komiske:2018cqr,Qu:2019gqs}. We generate a new sample following the process and jet selection of the public quark/gluon dataset~\cite{Komiske2019QGData}. The hard channels are $qg\to Zq$ and $q\bar q\to Zg$ in $pp$ collisions at $\sqrt{s}=14$~TeV, with $Z\to\nu\bar\nu$ and $q=u,d,s$ or their antiparticles for the quark-jet class. Generation uses \textsc{Pythia}~8.316, starting from the Monash 2013 tune~\cite{Skands:2014pea}, with initial- and final-state radiation, multiparton interactions (MPI), and hadronization enabled. Final-state particles other than neutrinos are clustered with anti-$k_T$, $R=0.4$, using the FastJet \texttt{fjcore}~3.3.0 package~\cite{Cacciari:2011ma}. We retain at most one jet per event: the highest-$p_T$ jet satisfying $500\leq p_T\leq550$~GeV and $|y|<1.7$. 
No detector simulation is applied. 

\begin{figure}[!t]
 \centering
 \includegraphics[width=\textwidth]{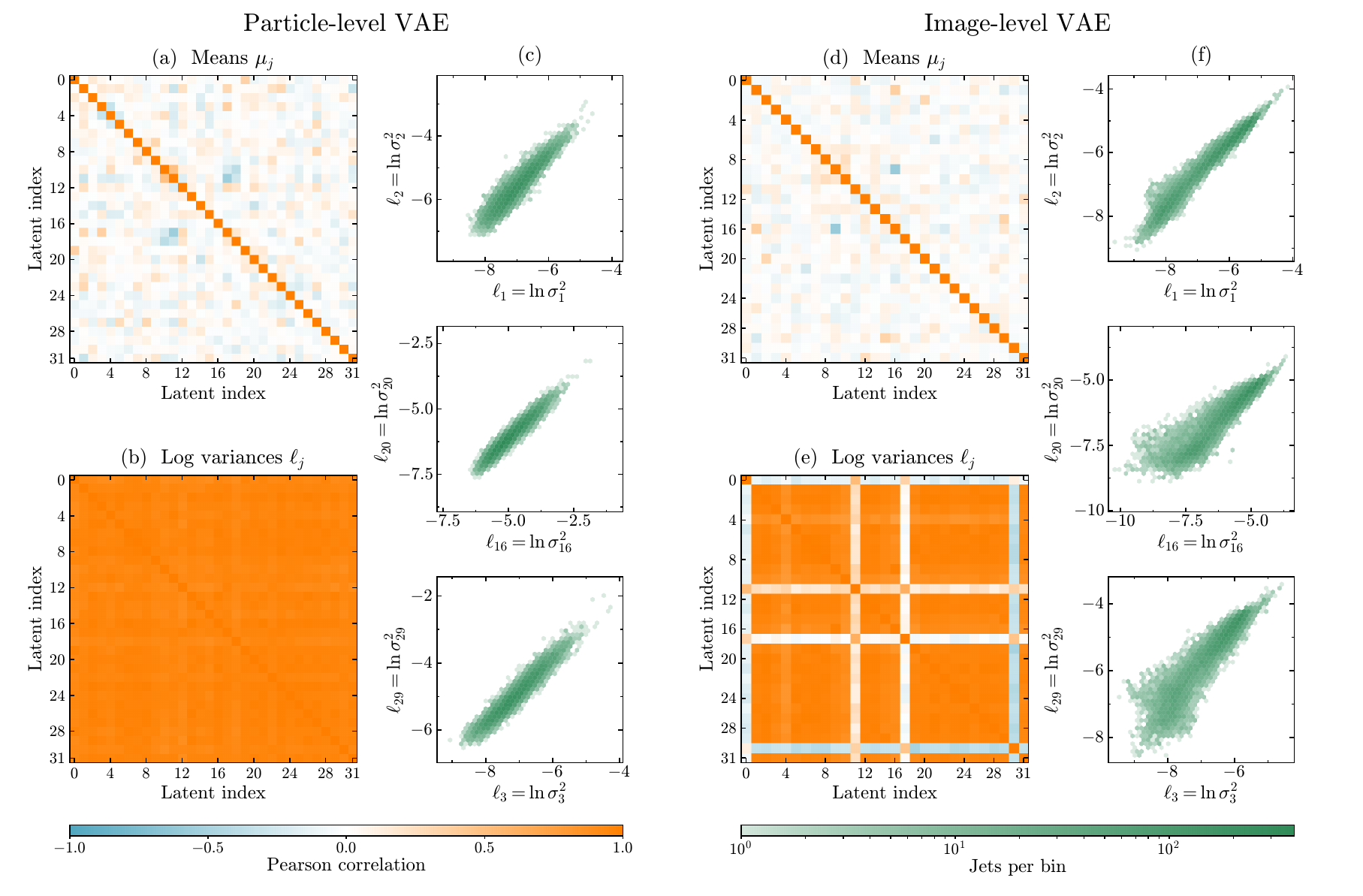}
 \caption{Posterior structure for $Z+$jet production in the particle-level VAE with $\beta=10^{-5}$ \textbf{(left)} and the image-level VAE with $\beta=2\times10^{-6}$ \textbf{(right)}. Panels \textbf{(a,d)} show the Pearson correlation matrices of the posterior means $\mu_j$, and panels \textbf{(b,e)} show those of the log variances $\ell_j=\ln\sigma_j^2$, restricted to the 32 and 29 active dimensions of the respective models as defined by~\cref{eq:activity}. Panels \textbf{(c)} and \textbf{(f)} each show the 2D joint distributions of log variances for three pairs of collective latent dimensions. All panels use the same mixed sample of 5000 quark jets and 5000 gluon jets.}
 \label{fig:zjet_posterior_structure}
\end{figure}

For the posterior-structure comparison, we use the same dijet-trained checkpoints as in \cref{fig:collective}, with $\beta=10^{-5}$ for the particle-level VAE and $\beta=2\times10^{-6}$ for the image-level VAE. Both models are evaluated on the same 5000 quark jets and 5000 gluon jets drawn from the default $Z+$jet sample, with the same preprocessing and no PID input. There are 32 and 29 active latent dimensions in the particle-level and image-level models, respectively. \cref{fig:zjet_posterior_structure}(a,d) shows weak correlations among the mean features $\mu_j$, with median off-diagonal $|r|=0.050$ and $0.046$, respectively. The corresponding log variances $\ell_j=\ln\sigma_j^2$ are strongly correlated in panels (b,e), with median off-diagonal $r=0.948$ and $0.929$. Panels (c,f) show the joint distributions for the same three pairs of collective latent dimensions in both models, also used in \cref{fig:collective}; their narrow, approximately linear structure illustrates this collective behavior.

The approximate common scaling also persists in this sample. We retain all 32 active particle-level dimensions and the same 25 collective image-level dimensions as in the dijet analysis. For both the dijet and $Z+$jet analyses, the coefficients $b_j$ are obtained separately for each model from the leading principal component of the centered $\ell_j$ values, without rescaling their individual variances, and normalized to unit mean. For the $Z+$jet sample, the first component explains $94.7\%$ and $94.2\%$ of the total variance in the respective retained sets. The standard deviations of the normalized $b_j$ are $0.0643$ and $0.0862$, respectively. These small spreads support the approximate common rescaling of the widths in \cref{eq:width_scaling}, with approximately invariant width ratios along the dominant collective direction.

For the parameter-response study below, we use the image-level checkpoint with $\beta=2\times10^{-6}$ and the same particle-level checkpoint.
We next vary the same eight generator parameters as in the main text by $-20\%$, $-10\%$, $0\%$, $+10\%$, and $+20\%$ relative to their default values.
For the $Z+$jet $\as$ direction, the coupling inputs for final-state radiation (\texttt{TimeShower}), initial-state radiation (\texttt{SpaceShower}), the hard process (\texttt{SigmaProcess}), and MPI (\texttt{Multiparton\allowbreak Interactions}) receive the same fractional change. The default \texttt{alphaSvalue} is 0.1365 for each shower and 0.130 for the hard process and MPI, giving ranges 0.1092--0.1638 and 0.104--0.156, respectively. These inputs override the corresponding tune values. The PDF set and its internal coupling are kept fixed. The remaining seven directions are varied one at a time, with the definitions, defaults, and ranges in Appendix~\ref{app:analysis} and \cref{tab:ranges}. Responses are measured relative to the default $Z+$jet sample, using the fixed encoders and the linear-response fit of \cref{eq:response}.

\begin{figure}[htbp]
 \centering
 \includegraphics[width=0.95\textwidth]{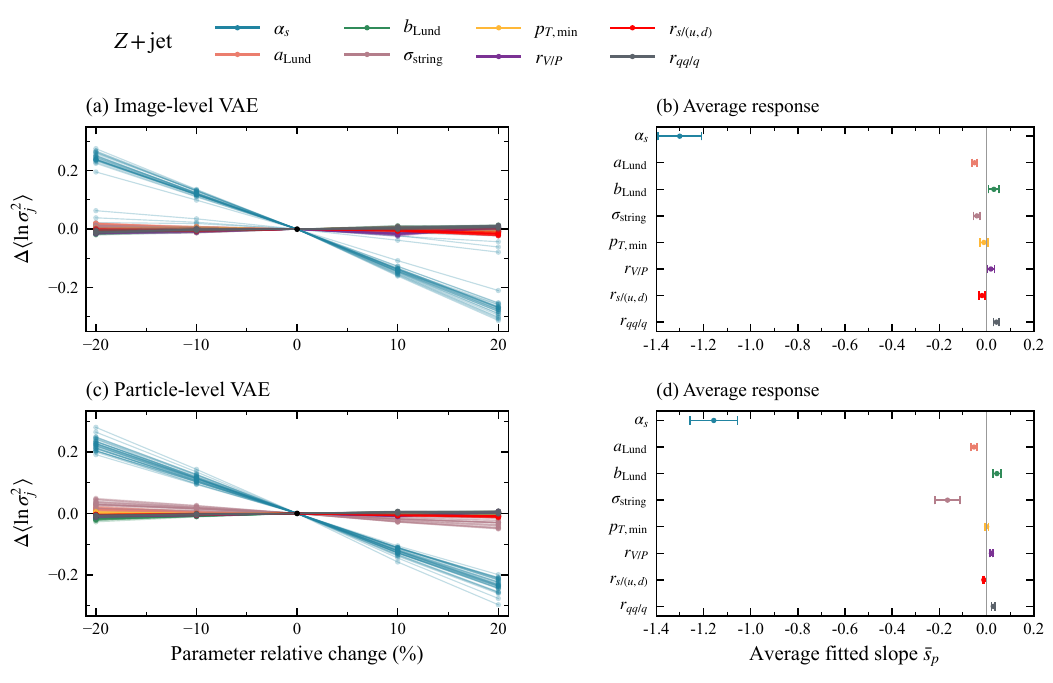}
 \caption{Collective parameter responses in the image-level VAE (top row) and particle-level VAE (bottom row) for $Z+$jet samples containing 10\,000 quark jets and 10\,000 gluon jets per setting. \textbf{(a,c)} Each thin curve connects the five measured shifts for one active dimension and one generator direction. All 29 active image-level dimensions and all 32 active particle-level dimensions are shown. \textbf{(b,d)} Small dots give the average fitted slopes $s_{j,p}$ over the active dimensions. Horizontal error bars show the standard deviation of the fitted slope across these dimensions. Four active dimensions in the image-level VAE are clear outliers, exhibiting much weaker responses to $\as$ than the others. To focus on the collective behavior, these four dimensions are excluded from the $\as$ average and standard deviation in (b).}
 \label{fig:zjet_all_active_response}
\end{figure}

\cref{fig:zjet_all_active_response} shows a strong, approximately linear $\as$ response in both representations. Among the eight directions, $\as$ gives the largest absolute fitted slope in all 32 active particle-level dimensions and most of the 29 active image-level dimensions.
Four active latent dimensions in the image-level model are clear outliers, exhibiting much weaker responses to $\as$ than the others. To focus on the collective behavior, we exclude these four dimensions from the image-level slope summary. The average slopes are $-1.301$ and $-1.157$, with standard deviations of 0.092 and 0.101 over the retained 25 image-level dimensions and all 32 particle-level dimensions, respectively. The other parameter directions give much smaller collective responses. Thus, the strongly correlated widths, approximate common scaling, and pronounced coupling response found in the dijet sample study also occur for this $Z+$jet sample with the trained models held fixed.

\section{Multiplicity dependence and a phase-space comparison}
\label{app:rambo}

To test whether the collective variation of $\ln\sigma_j^2$ reflects only particle multiplicity, we compare QCD jets with events generated using the RAndom Momenta Beautifully Organized (RAMBO) algorithm~\cite{Kleiss:1985gy}, which uniformly samples Lorentz-invariant phase space for massless final-state particles. At each fixed multiplicity, these events can be interpreted as the decay of a massive scalar into identical massless scalars with a constant matrix element. The RAMBO events have an invariant mass of 20~GeV and are boosted to a total momentum of 200~GeV along the $x$ axis. We analyze both samples using the same trained particle-level VAE, taking the common variable $u(x)$ defined in \cref{eq:factor} as a diagnostic, because it mainly governs the information in $\sigma$-space.

\begin{figure}[H]
 \centering
 \includegraphics[width=\textwidth]{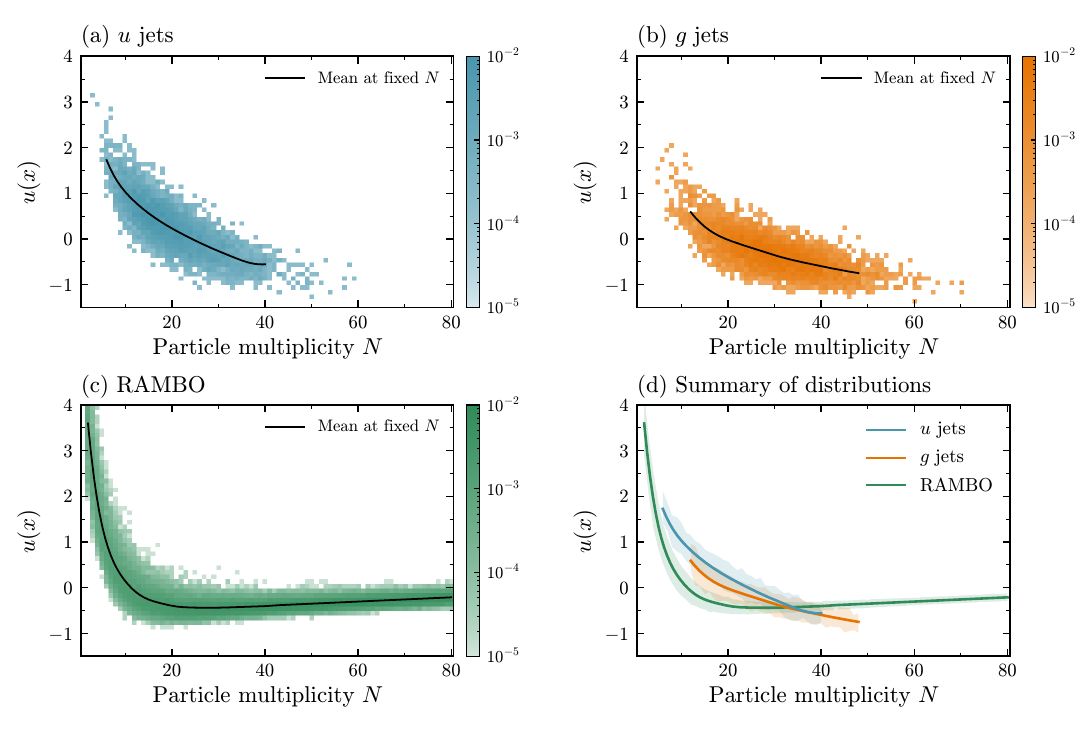}
 \caption{Common posterior-width variable $u(x)$ versus particle multiplicity $N$ for \textbf{(a)} up-quark jets, \textbf{(b)} gluon jets, and \textbf{(c)} RAMBO events. Black solid lines show the mean at fixed $N$. Panel \textbf{(d)} compares these means, with shaded bands showing the central 68\% of events at fixed $N$, rather than uncertainties on the means.}
 \label{fig:rambo_multiplicity}
\end{figure}

\cref{fig:rambo_multiplicity} shows that $u(x)$ depends on particle multiplicity $N$ in both QCD jets and RAMBO events. Their distributions, however, are different. If the posterior widths encoded only the particle count, the two samples would fall into the same region in the 2D plot of $u(x)$ vs. $N$. Instead, the conditional means differ, and the central 68\% bands are separated over part of the multiplicity range, as illustrated in panel (d). Thus the common variable cannot be identified with particle multiplicity alone.

Changing $\as$ also changes the particle multiplicity of QCD jets. We therefore ask whether the associated change in multiplicity can reproduce the response of $\langle\ln\sigma_j^2\rangle$. We construct RAMBO ensembles matching the full multiplicity distributions of the five QCD samples in the $\as$ scan. This comparison includes the changes in the shape of the multiplicity distribution as $\as$ is varied.

\begin{figure}[H]
 \centering
 \includegraphics[width=0.51\textwidth]{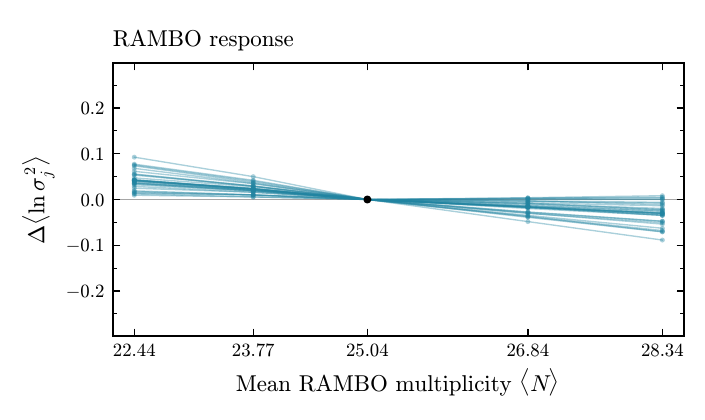}
 \caption{Responses in $\langle\ln\sigma_j^2\rangle$ for RAMBO ensembles matched to the full multiplicity distributions of the five QCD coupling samples, shown against the mean particle multiplicity. Each curve corresponds to one active latent dimension. Shifts are measured relative to the RAMBO ensemble matched to the default QCD sample ($\Delta\as=0$).}
 \label{fig:rambo_response}
\end{figure}

As shown in \cref{fig:rambo_response}, the RAMBO responses remain much smaller than the QCD coupling response in \cref{fig:parameters}. Particle multiplicity is therefore relevant to the learned widths, but these comparisons show that the widths also reflect other aspects of the particle configuration. If $\sigma$-space were mostly determined by the number of jet particles, we would expect a response comparable to that obtained by varying $\as$.

\section{Correlations with standard jet observables}
\label{app:observables}

\begin{figure}[ht]
 \centering
 \includegraphics[height=0.773818\textheight,width=\textwidth,keepaspectratio]{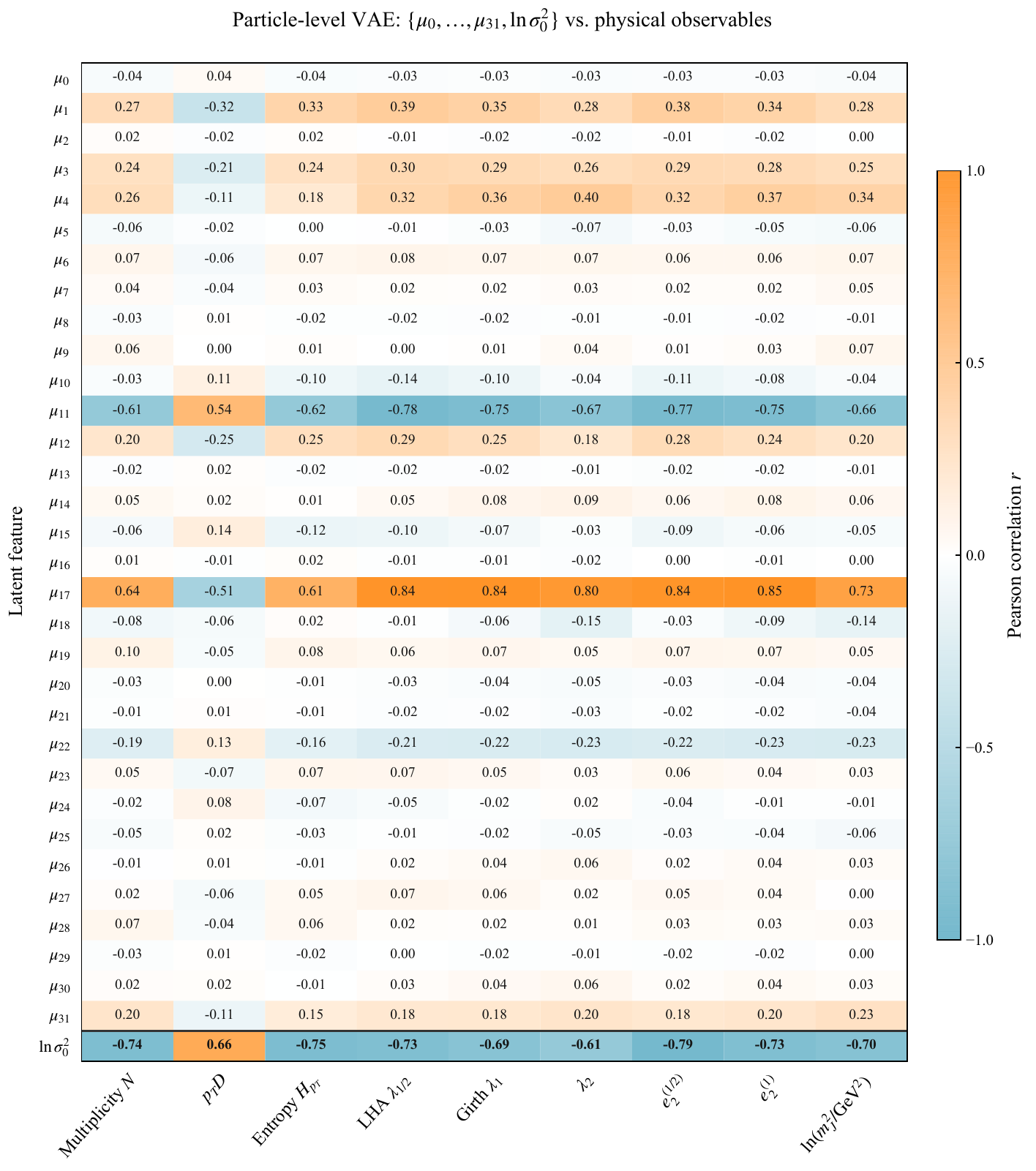}
 \caption{Signed Pearson correlations of the 32 active particle-level VAE mean coordinates $\mu_0,\ldots,\mu_{31}$ and the log variance of the first active dimension, $\ell_0=\ln\sigma_0^2$, with the nine jet observables defined in \cref{tab:observable_definitions}. The $33\times9$ matrix uses the same 5000 quark jets and 5000 gluon jets as \cref{fig:collective}. Observables use all positive-$p_T$ constituents and the angular reference scale $R_{\mathrm{ref}}=0.4$.}
 \label{fig:observable_correlations}
\end{figure}

Correlations with familiar jet observables help characterize the physical information associated with the learned posterior parameters. \cref{fig:observable_correlations} shows the Pearson correlations of the particle-level VAE mean coordinates $\mu_j$ and one $\sigma$-coordinate $\ell_0=\ln\sigma_0^2$ with jet observables selected for their strong correlations with $\ell_0$. These observables characterize constituent multiplicity, transverse-momentum sharing, angular structure and jet mass. The evaluation uses 5000 quark jets and 5000 gluon jets from the dijet sample.

The nine observables in \cref{fig:observable_correlations} are defined in \cref{tab:observable_definitions}. LHA denotes the Les Houches angularity. All observables are dimensionless. We use $R_{\mathrm{ref}}=0.4$ and define
\begin{equation}
 z_i=\frac{p_{T,i}}{\sum_k p_{T,k}},\quad
 \theta_i=\frac{\Delta R_{iJ}}{R_{\mathrm{ref}}},\quad
 \theta_{ij}=\frac{\Delta R_{ij}}{R_{\mathrm{ref}}},\quad
 \Delta R_{ab}=\sqrt{(\eta_a-\eta_b)^2+
 \operatorname{wrap}(\phi_a-\phi_b)^2},
 \label{eq:observable_coordinates}
\end{equation}
where $\operatorname{wrap}(\Delta\phi)\in[-\pi,\pi)$. The jet axis is obtained from the summed constituent four-momentum $p_J^\mu=\sum_i p_i^\mu$:
\begin{equation}
 p_{T,J}=\sqrt{p_{x,J}^2+p_{y,J}^2},\quad
 \eta_J=\operatorname{arsinh}\!\left(\frac{p_{z,J}}{p_{T,J}}\right),\quad
 \phi_J=\operatorname{atan2}(p_{y,J},p_{x,J}),\quad
 m_J^2=E_J^2-|\boldsymbol p_J|^2.
 \label{eq:observable_axis}
\end{equation}
The scalar sum $\sum_i p_{T,i}$ defines the momentum fractions $z_i$, while the summed four-momentum determines the jet axis and mass. The angular normalization scale $R_{\mathrm{ref}}$ is chosen to match the jet-clustering radius $R=0.4$, and each pair in the energy correlators is counted once.

In \cref{fig:observable_correlations}, the log variance is negatively correlated with multiplicity ($r=-0.74$), transverse-momentum entropy ($r=-0.75$), the two-point energy correlator $e_2^{(1/2)}$ ($r=-0.79$), and $\ln(m_J^2/\mathrm{GeV}^2)$ ($r=-0.70$), and positively correlated with the momentum concentration $p_TD$ ($r=0.66$). Several mean coordinates, particularly $\mu_{11}$ and $\mu_{17}$, also correlate strongly with these observables.

The correlation pattern associates smaller posterior widths with higher constituent multiplicity, more broadly shared transverse momentum, and larger angular extent and jet mass. These associations connect the learned widths to familiar features of the radiation pattern within a jet. They are consistent with the sensitivity of the collective posterior widths to $\as$ observed in the parameter scan.

\begin{table}[htbp]
\centering
\small
\renewcommand{\arraystretch}{1.5}
\setlength{\tabcolsep}{4pt}
\caption{Definitions of the observables in \cref{fig:observable_correlations}. The angularities use the summed-four-momentum jet axis.}
\label{tab:observable_definitions}
\begin{tabular}{r p{0.18\textwidth} p{0.36\textwidth} p{0.35\textwidth}}
\toprule
\# & Observable & Definition & Interpretation \\
\midrule
1 & Multiplicity $N$ & $\displaystyle\sum_{i:p_{T,i}>0}1$ & Number of constituents. \\[5pt]
2 & $p_TD$ & $\displaystyle\sqrt{\sum_i z_i^2}$ & Transverse-momentum concentration. \\[5pt]
3 & Entropy $H_{p_T}$ & $\displaystyle-\sum_i z_i\ln z_i$ & Spread of momentum fractions. \\[5pt]
4 & LHA $\lambda_{1/2}$ & $\displaystyle\sum_i z_i\theta_i^{1/2}$ & Angularity with exponent $1/2$. \\[5pt]
5 & Normalized girth $\lambda_1$ & $\displaystyle\sum_i z_i\theta_i$ & Momentum-weighted radial extent. \\[5pt]
6 & $\lambda_2$ & $\displaystyle\sum_i z_i\theta_i^2$ & Second radial moment. \\[5pt]
7 & $e_2^{(1/2)}$ & $\displaystyle\sum_{i<j}z_i z_j\theta_{ij}^{1/2}$ & Two-point energy correlator, exponent $1/2$. \\[5pt]
8 & $e_2^{(1)}$ & $\displaystyle\sum_{i<j}z_i z_j\theta_{ij}$ & Two-point energy correlator, exponent $1$. \\[5pt]
9 & $\ln(m_J^2/\mathrm{GeV}^2)$ & $\displaystyle\ln\!\left(\frac{m_J^2}{\mathrm{GeV}^2}\right)$ & Logarithm of jet mass squared. \\[5pt]
\bottomrule
\end{tabular}
\end{table}

\section{Comparison of posterior-width structure across datasets}
\label{app:datasets}

The jet comparisons in \cref{tab:dataset_widths} motivate our choices of the KL weight $\beta$. For the image-level VAE, $\beta=2\times10^{-6}$ retains 29 of the 32 latent dimensions as active, whereas increasing $\beta$ to $10^{-5}$ and $10^{-4}$ leaves only five and three active dimensions. We therefore choose $\beta=2\times10^{-6}$ to retain a richer representation of jet structure for reconstruction. For the particle-level VAE, $\beta=10^{-5}$ and $10^{-4}$ retain 32 and 31 active dimensions, so both choices preserve a large number of active features. At $\beta=10^{-3}$, however, only four dimensions remain active. We eventually choose $\beta=10^{-5}$ for the particle-level analysis.

To quantify the correlation structure, we use the Pearson correlation matrices $C^{\mu}$ and $C^{\ell}$ of the posterior means $\mu_j$ and log variances $\ell_j=\ln\sigma_j^2$, evaluated across inputs on the $N$ active latent dimensions. For either matrix, the effective rank is defined by the participation ratio of its eigenvalues $\lambda_1,\dots,\lambda_N\geq0$,
\begin{equation}
 r_{\mathrm{eff}}(C^{a})=\frac{\left(\sum_{k}\lambda_k\right)^2}{\sum_{k}\lambda_k^2},
 \qquad a=\mu,\ell \, ,
 \label{eq:app_rank}
\end{equation}
with $1\leq r_{\mathrm{eff}}(C^{a})\leq N$. This quantity characterizes how broadly the variation is distributed among the correlation eigenmodes. A value close to one indicates a single dominant mode of variation across inputs, while larger values indicate several appreciable, mutually uncorrelated modes. It equals one when all active features are perfectly correlated or anticorrelated, and $N$ when they are all uncorrelated. It can be smaller than $N$ even when the correlation matrix has full algebraic rank. In \cref{tab:dataset_widths,tab:dataset_widths_extra}, the posterior means retain many appreciable correlation modes, although accumulated weak correlations can reduce $r_{\mathrm{eff}}(C^{\mu})$ below the number of active coordinates.

The posterior log variances show a different pattern. Their effective ranks $r_{\mathrm{eff}}(C^{\ell})$ are substantially smaller than $r_{\mathrm{eff}}(C^{\mu})$ and need not follow the number of active dimensions. The non-jet benchmarks in \cref{tab:dataset_widths_extra} also exhibit correlated log variances, with effective ranks that depend on the dataset and training settings. The jet models exhibit a dominant collective mode across the tested regularization strengths, while retaining a substantially richer structure in their posterior means. This separation is the striking feature: a jet representation can retain many distinct mean-feature directions while their variances vary predominantly through one common degree of freedom. 
Our controlled coupling-response study connects this collective variance structure to the underlying QCD dynamics.

Multiple appreciable correlation modes can arise when the posterior widths of different features exhibit distinct patterns of variation across inputs. For example, if different latent features encode color and spatial structure, their corresponding widths need not vary together from one image to another. Such differences can lead to a larger effective rank. For jets, the near-one effective rank indicates that the widths vary predominantly through a common mode, even though the posterior means retain many effective degrees of freedom.

\paragraph{Non-jet reconstruction loss.}
The non-jet image benchmarks all use the binary-cross-entropy (BCE) reconstruction cost,
\begin{equation}
 L_{\mathrm{reco}}^{\mathrm{non\text{-}jet}}
 =-\sum_{p=1}^{P}
 \left[x_p\ln\tilde{x}_p
 +(1-x_p)\ln(1-\tilde{x}_p)\right],
 \label{eq:nonjet-bce}
\end{equation}
where $P=C\times H\times W$ counts all pixels and channels, $x_p\in[0,1]$ is the target intensity, and $\tilde{x}_p\in(0,1)$ is the decoder output. During training, this per-image reconstruction loss is averaged over each batch of 64 images. Contributions are summed over pixels and channels without division by $P$. This reconstruction term is combined with the Gaussian KL term in \cref{eq:loss}.
For the binary dSprites images, $x_p\in\{0,1\}$, BCE is the negative log-likelihood of a conditionally factorized Bernoulli decoder, with $\tilde{x}_p$ giving the probability that a pixel equals one. For grayscale or color images with intermediate intensities, the same expression is used as a reconstruction cost with continuous target values in $[0,1]$. 

The numerical value of $\beta$ should not be interpreted as a common resolution scale across datasets or models. Its meaning depends on the normalization of the reconstruction objective. Since the KL divergence is dimensionless, $\beta$ has the same units as $L_{\mathrm{reco}}$ in \cref{eq:loss}. The two jet reconstruction losses are dimensionless: the image loss uses normalized transverse-momentum fractions, while the transport loss uses normalized constituent weights and angular distances. The BCE cost in \cref{eq:nonjet-bce} is also dimensionless. Thus $\beta$ is dimensionless in our implementation, although an unnormalized energy-based cost would require a coefficient with the corresponding units. Rescaling $L_{\mathrm{reco}}\to cL_{\mathrm{reco}}$ and $\beta\to c\beta$ together, with $c>0$, leaves the objective's minimizers unchanged. For a $32\times32$ non-jet image, each pixel contributes a reconstruction loss of $\mathcal{O}(1)$, giving a total loss of $\mathcal{O}(10^3)$ per sample. Choosing $\beta\sim\mathcal{O}(1)$ under this convention is equivalent to rescaling the reconstruction loss to $\mathcal{O}(1)$ and using $\beta\sim\mathcal{O}(10^{-3})$. At fixed normalization, varying $\beta$ changes the reconstruction-information tradeoff. For jets, this tradeoff governs the precision with which energy flow and substructure are retained under the chosen loss.

We report $\beta/\langle L_{\mathrm{reco}}\rangle$ in \cref{tab:dataset_widths,tab:dataset_widths_extra} to indicate the numerical scale of the KL coefficient relative to the mean reconstruction cost under each loss convention. This ratio is unchanged by a common rescaling of $L_{\mathrm{reco}}$ and $\beta$. If the reconstruction is sufficiently accurate, the posterior variances may depend mainly on how sharply the reconstruction loss increases when the latent coordinates are slightly displaced from their posterior means. A curvature-based comparison would require an additional evaluation of the reconstruction Hessian and its variation across inputs and latent directions, which is beyond the scope of this study.

\begin{table}[htbp]
\centering
\small
\setlength{\tabcolsep}{5pt}
\caption{Posterior means $\mu_j$ and log variances $\ell_j=\ln\sigma_j^2$ in jet samples, comparing the image-level CNN and particle-level Transformer (Trf) representations. The models are trained on the dijet sample and held fixed for the evaluations shown here; $\beta$ denotes the final effective KL coefficient. The dijet CNN row at $\beta=2\times10^{-6}$ and Transformer row at $\beta=10^{-5}$ are highlighted in bold. ``Active'' lists the number of active latent dimensions (defined by \cref{eq:activity}) out of the total. The effective ranks are participation ratios of the correlation matrices $C^{a}$ or covariance matrices $\Sigma^{a}$, with $a=\mu,\ell$, as defined in the text. The correlation columns are defined in \cref{eq:app_corr}.}
\label{tab:dataset_widths}
\begin{tabular}{lrrcrrrrrrr}
\toprule
Dataset & $\beta$ & $\beta/\langle L_{\mathrm{reco}}\rangle$ & Model & Active & Corr.\ $\ell$ & $r_{\mathrm{eff}}(C^{\ell})$ & $r_{\mathrm{eff}}(\Sigma^{\ell})$ & $|\mathrm{Corr.}\ \mu|$ & $r_{\mathrm{eff}}(C^{\mu})$ & $r_{\mathrm{eff}}(\Sigma^{\mu})$ \\
\midrule
Jets (dijet) & $10^{-6}$ & $2.6\times10^{-4}$ & CNN & 30/32 & $+0.72$ & $1.35$ & $1.14$ & $0.07$ & $23.4$ & $19.7$ \\
\textbf{Jets (dijet)} & $\bm{2\times10^{-6}}$ & $\bm{4.7\times10^{-4}}$ & \textbf{CNN} & \textbf{29/32} & $\bm{+0.67}$ & $\bm{1.52}$ & $\bm{1.25}$ & $\bm{0.05}$ & $\bm{25.7}$ & $\bm{23.9}$ \\
Jets (dijet) & $3\times10^{-6}$ & $6.2\times10^{-4}$ & CNN & 23/32 & $+0.68$ & $1.50$ & $1.23$ & $0.04$ & $21.6$ & $20.9$ \\
Jets (dijet) & $4\times10^{-6}$ & $8.1\times10^{-4}$ & CNN & 23/32 & $+0.70$ & $1.50$ & $1.15$ & $0.04$ & $21.4$ & $20.7$ \\
Jets (dijet) & $5\times10^{-6}$ & $7.9\times10^{-4}$ & CNN & 16/32 & $+0.68$ & $1.73$ & $1.29$ & $0.06$ & $14.3$ & $13.8$ \\
Jets (dijet) & $10^{-5}$ & $5.7\times10^{-4}$ & CNN & 5/32 & $+0.93$ & $1.12$ & $1.08$ & $0.09$ & $4.8$ & $4.5$ \\
Jets (dijet) & $10^{-4}$ & $4.4\times10^{-3}$ & CNN & 3/32 & $+0.89$ & $1.17$ & $1.16$ & $0.05$ & $3.0$ & $3.0$ \\
\addlinespace
\textbf{Jets (dijet)} & $\bm{10^{-5}}$ & $\bm{3.2\times10^{-4}}$ & \textbf{Trf} & \textbf{32/32} & $\bm{+0.93}$ & $\bm{1.15}$ & $\bm{1.15}$ & $\bm{0.05}$ & $\bm{27.5}$ & $\bm{26.3}$ \\
Jets (dijet) & $10^{-4}$ & $4.3\times10^{-3}$ & Trf & 31/32 & $+0.80$ & $1.47$ & $1.66$ & $0.02$ & $30.6$ & $30.5$ \\
Jets (dijet) & $2\times10^{-4}$ & $8.0\times10^{-3}$ & Trf & 30/32 & $+0.78$ & $1.55$ & $1.61$ & $0.02$ & $29.6$ & $29.5$ \\
Jets (dijet) & $3\times10^{-4}$ & $9.9\times10^{-3}$ & Trf & 22/32 & $+0.75$ & $1.69$ & $1.51$ & $0.01$ & $21.9$ & $21.8$ \\
Jets (dijet) & $4\times10^{-4}$ & $1.3\times10^{-2}$ & Trf & 22/32 & $+0.75$ & $1.67$ & $1.49$ & $0.01$ & $21.9$ & $21.7$ \\
Jets (dijet) & $5\times10^{-4}$ & $1.4\times10^{-2}$ & Trf & 18/32 & $+0.73$ & $1.72$ & $1.54$ & $0.01$ & $17.9$ & $17.6$ \\
Jets (dijet) & $10^{-3}$ & $1.5\times10^{-2}$ & Trf & 4/32 & $+0.84$ & $1.28$ & $1.28$ & $0.07$ & $3.9$ & $3.9$ \\
\addlinespace
Jets ($Z+$jet) & $2\times10^{-6}$ & $8.2\times10^{-4}$ & CNN & 29/32 & $+0.68$ & $1.47$ & $1.21$ & $0.06$ & $24.3$ & $18.8$ \\
Jets ($Z+$jet) & $10^{-5}$ & $4.7\times10^{-4}$ & Trf & 32/32 & $+0.95$ & $1.11$ & $1.11$ & $0.07$ & $24.0$ & $18.4$ \\
\bottomrule
\end{tabular}
\end{table}

\begin{table}[htbp]
\centering
\small
\setlength{\tabcolsep}{5pt}
\caption{Posterior means $\mu_j$ and log variances $\ell_j=\ln\sigma_j^2$ in non-jet samples, with the same column definitions as \cref{tab:dataset_widths}. Models labeled ``This work'' were trained by ourselves, whereas those labeled ``Pretrained'' use externally provided pretrained checkpoints. All values use the original evaluation samples and active-dimension criterion.}
\label{tab:dataset_widths_extra}
\begin{tabular}{lrrcrrrrrrr}
\toprule
Dataset & $\beta$ & $\beta/\langle L_{\mathrm{reco}}\rangle$ & Model source & Active & Corr.\ $\ell$ & $r_{\mathrm{eff}}(C^{\ell})$ & $r_{\mathrm{eff}}(\Sigma^{\ell})$ & $|\mathrm{Corr.}\ \mu|$ & $r_{\mathrm{eff}}(C^{\mu})$ & $r_{\mathrm{eff}}(\Sigma^{\mu})$ \\
\midrule
MNIST & $4$ & $3.2\times10^{-2}$ & Pretrained & 8/10 & $+0.26$ & $3.89$ & $2.95$ & $0.03$ & $7.9$ & $7.7$ \\
MNIST & $4$ & $3.1\times10^{-2}$ & This work & 10/10 & $+0.39$ & $3.06$ & $2.44$ & $0.03$ & $9.9$ & $9.4$ \\
MNIST & $1$ & $8.5\times10^{-3}$ & Pretrained & 10/10 & $+0.58$ & $2.09$ & $1.44$ & $0.03$ & $9.9$ & $9.8$ \\
MNIST & $1$ & $8.3\times10^{-3}$ & This work & 10/10 & $+0.62$ & $1.82$ & $1.51$ & $0.05$ & $9.6$ & $9.5$ \\
MNIST & $0.1$ & $8.4\times10^{-4}$ & This work & 10/10 & $+0.76$ & $1.57$ & $1.45$ & $0.11$ & $8.7$ & $8.5$ \\
MNIST & $4$ & $3.1\times10^{-2}$ & This work & 8/32 & $+0.39$ & $2.73$ & $2.31$ & $0.04$ & $7.9$ & $7.8$ \\
MNIST & $1$ & $8.8\times10^{-3}$ & This work & 17/32 & $+0.49$ & $2.58$ & $2.06$ & $0.04$ & $16.4$ & $15.7$ \\
MNIST & $0.1$ & $9.5\times10^{-4}$ & This work & 32/32 & $+0.80$ & $1.47$ & $1.33$ & $0.07$ & $26.2$ & $25.3$ \\
MNIST & $4$ & $3.1\times10^{-2}$ & This work & 8/128 & $+0.31$ & $3.41$ & $2.55$ & $0.03$ & $7.9$ & $7.7$ \\
MNIST & $1$ & $8.8\times10^{-3}$ & This work & 17/128 & $+0.65$ & $2.06$ & $1.86$ & $0.04$ & $16.3$ & $15.6$ \\
MNIST & $0.1$ & $9.7\times10^{-4}$ & This work & 45/128 & $+0.81$ & $1.48$ & $1.35$ & $0.04$ & $39.7$ & $37.4$ \\
\addlinespace
FashionMNIST & $4$ & $1.3\times10^{-2}$ & This work & 5/10 & $+0.18$ & $3.92$ & $3.49$ & $0.03$ & $5.0$ & $4.7$ \\
FashionMNIST & $1$ & $3.3\times10^{-3}$ & This work & 10/10 & $+0.34$ & $3.73$ & $3.37$ & $0.04$ & $9.7$ & $9.5$ \\
FashionMNIST & $0.1$ & $3.3\times10^{-4}$ & This work & 10/10 & $+0.52$ & $2.56$ & $2.65$ & $0.13$ & $8.1$ & $7.9$ \\
FashionMNIST & $4$ & $1.3\times10^{-2}$ & This work & 5/32 & $+0.09$ & $3.34$ & $2.87$ & $0.03$ & $4.9$ & $4.7$ \\
FashionMNIST & $1$ & $3.3\times10^{-3}$ & This work & 12/32 & $+0.18$ & $3.83$ & $3.47$ & $0.04$ & $11.7$ & $11.1$ \\
FashionMNIST & $0.1$ & $3.4\times10^{-4}$ & This work & 32/32 & $+0.39$ & $3.12$ & $2.84$ & $0.05$ & $27.8$ & $25.6$ \\
FashionMNIST & $4$ & $1.3\times10^{-2}$ & This work & 5/128 & $+0.10$ & $3.77$ & $3.25$ & $0.03$ & $5.0$ & $4.7$ \\
FashionMNIST & $1$ & $3.3\times10^{-3}$ & This work & 12/128 & $+0.27$ & $3.71$ & $3.39$ & $0.03$ & $11.8$ & $11.1$ \\
FashionMNIST & $0.1$ & $3.4\times10^{-4}$ & This work & 33/128 & $+0.31$ & $2.81$ & $2.57$ & $0.04$ & $29.9$ & $27.4$ \\
\addlinespace
CelebA & $10$ & $1.5\times10^{-3}$ & Pretrained & 10/10 & $+0.42$ & $2.63$ & $2.00$ & $0.02$ & $9.9$ & $9.9$ \\
CelebA & $1$ & $1.5\times10^{-4}$ & Pretrained & 10/10 & $+0.49$ & $2.82$ & $2.70$ & $0.04$ & $9.6$ & $9.5$ \\
CelebA & $0.1$ & $1.5\times10^{-5}$ & This work & 10/10 & $+0.78$ & $1.55$ & $1.56$ & $0.11$ & $8.3$ & $8.1$ \\
CelebA & $4$ & $6.4\times10^{-4}$ & This work & 32/32 & $+0.58$ & $2.40$ & $2.37$ & $0.02$ & $31.6$ & $31.1$ \\
CelebA & $1$ & $1.6\times10^{-4}$ & This work & 32/32 & $+0.74$ & $1.75$ & $1.68$ & $0.03$ & $30.6$ & $30.5$ \\
CelebA & $0.1$ & $1.6\times10^{-5}$ & This work & 32/32 & $+0.79$ & $1.58$ & $1.56$ & $0.09$ & $21.7$ & $19.7$ \\
CelebA & $4$ & $6.4\times10^{-4}$ & This work & 41/128 & $+0.57$ & $2.42$ & $2.58$ & $0.01$ & $40.5$ & $38.9$ \\
CelebA & $1$ & $1.6\times10^{-4}$ & This work & 71/128 & $+0.77$ & $1.65$ & $1.61$ & $0.02$ & $68.6$ & $67.5$ \\
CelebA & $0.1$ & $1.6\times10^{-5}$ & This work & 110/128 & $+0.80$ & $1.55$ & $1.52$ & $0.03$ & $92.8$ & $81.4$ \\
\addlinespace
dSprites & $4$ & $1.7\times10^{-1}$ & Pretrained & 8/10 & $+0.05$ & $6.12$ & $4.36$ & $0.02$ & $8.0$ & $7.6$ \\
dSprites & $1$ & $7.8\times10^{-2}$ & Pretrained & 10/10 & $+0.10$ & $5.20$ & $4.61$ & $0.03$ & $9.9$ & $9.7$ \\
dSprites & $0.1$ & $1.1\times10^{-2}$ & This work & 10/10 & $+0.17$ & $5.70$ & $5.81$ & $0.13$ & $7.9$ & $7.8$ \\
dSprites & $4$ & $1.7\times10^{-1}$ & This work & 9/32 & $+0.05$ & $5.68$ & $4.63$ & $0.03$ & $8.9$ & $8.6$ \\
dSprites & $1$ & $7.4\times10^{-2}$ & This work & 11/32 & $-0.02$ & $4.80$ & $3.19$ & $0.04$ & $10.7$ & $10.1$ \\
dSprites & $0.1$ & $1.3\times10^{-2}$ & This work & 19/32 & $+0.11$ & $6.34$ & $5.02$ & $0.04$ & $17.8$ & $16.3$ \\
dSprites & $4$ & $1.6\times10^{-1}$ & This work & 7/128 & $+0.05$ & $5.80$ & $4.97$ & $0.03$ & $6.9$ & $6.7$ \\
dSprites & $1$ & $7.9\times10^{-2}$ & This work & 11/128 & $+0.02$ & $5.75$ & $4.96$ & $0.03$ & $10.8$ & $10.4$ \\
dSprites & $0.1$ & $1.4\times10^{-2}$ & This work & 17/128 & $+0.14$ & $6.63$ & $5.03$ & $0.06$ & $14.9$ & $13.5$ \\
\bottomrule
\end{tabular}
\end{table}

\paragraph{Definitions of the table entries.}
All statistics are computed on a fixed evaluation sample of each dataset and only on the active latent dimensions. 
Let $\mathbb{E}_x$, $\operatorname{Var}_x$ and $\operatorname{Cov}_x$ denote the mean, variance and covariance over the samples $x$ of the evaluation set. A latent dimension $j$ is active when it satisfies \cref{eq:activity},
\begin{equation}
 \frac{\operatorname{Var}_x[\mu_j(x)]}{\mathbb{E}_x[\sigma_j^2(x)]}>1,
 \label{eq:app_active}
\end{equation}
which is also the criterion used for the non-jet models. We denote the set of active latent dimensions by $\mathcal{A}$, its size by $N$, and the total latent-space dimensionality by $d_z$. The column ``Active'' lists $N/d_z$. For a feature $a_j(x)$, which is either $\mu_j(x)$ or $\ell_j(x)=\ln\sigma_j^2(x)$, the Pearson correlation across the evaluation sample is
\begin{equation}
 \rho^{a}_{ij}=\frac{\operatorname{Cov}_x[a_i,a_j]}{\sqrt{\operatorname{Var}_x[a_i]\,\operatorname{Var}_x[a_j]}},\qquad i,j\in\mathcal{A},
 \label{eq:app_pearson}
\end{equation}
and $C^{a}=(\rho^{a}_{ij})_{i,j\in\mathcal{A}}$ denotes the resulting $N\times N$ correlation matrix, which has unit diagonal. The two correlation columns are averages over its off-diagonal elements,
\begin{equation}
 \text{Corr.}\ \ell=\frac{1}{N(N-1)}\sum_{\substack{i,j\in\mathcal{A}\\ i\neq j}}\rho^{\ell}_{ij},
 \qquad
 |\text{Corr.}\ \mu|=\frac{1}{N(N-1)}\sum_{\substack{i,j\in\mathcal{A}\\ i\neq j}}\bigl|\rho^{\mu}_{ij}\bigr|.
 \label{eq:app_corr}
\end{equation}
We average the absolute correlations among $\mu_j$ to avoid cancellations between positive and negative contributions. Even with this choice, the average absolute correlation among the posterior means remains much smaller than the signed average correlation among the log variances in the jet samples. 

Since $\sum_k\lambda_k=\operatorname{tr}C^{a}=N$ and $\sum_k\lambda_k^2=\operatorname{tr}\bigl[(C^{a})^2\bigr]=\sum_{i,j\in\mathcal{A}}(\rho^{a}_{ij})^2$, the effective rank in \cref{eq:app_rank} can equivalently be evaluated directly from the pairwise correlations,
\begin{equation}
 r_{\mathrm{eff}}(C^{a})=\frac{N^2}{\sum_{i,j\in\mathcal{A}}\left(\rho^{a}_{ij}\right)^2},
 \qquad a=\mu,\ell \, .
 \label{eq:app_rank_corr}
\end{equation}

The columns $r_{\mathrm{eff}}(\Sigma^{\ell})$ and $r_{\mathrm{eff}}(\Sigma^{\mu})$ use instead the covariance matrix $\Sigma^{a}=(\operatorname{Cov}_x[a_i,a_j])_{i,j\in\mathcal{A}}$ of the active dimensions, i.e.\ the features are centered but not divided by their standard deviations, so that dimensions with a larger variance carry more weight. Their effective rank is $r_{\mathrm{eff}}(\Sigma^{a})=\bigl(\operatorname{tr}\Sigma^{a}\bigr)^2/\operatorname{tr}\bigl[(\Sigma^{a})^2\bigr]$, with the eigenvalues of $\Sigma^{a}$ in place of those of $C^{a}$.

\end{document}